\documentclass[useAMS,usenatbib,authoryear]{mn2e}

\usepackage[utf8x]{inputenc}
\usepackage{graphicx} 
\usepackage{dcolumn}  
\usepackage{bm}       
\usepackage{amssymb,amsmath}  
\usepackage{color}
\usepackage{float}
\usepackage{natbib}
\def\Mpc{\, h^{-1} \, {\rm Mpc}}
\def\Gpc{\, h^{-1} \, {\rm Gpc}}

\def\Mo{\, h^{-1} \, {\rm M_{\odot}}}

\begin{document}

\title[A new method to measure galaxy bias by combining the density and weak lensing fields]{A new method to measure galaxy bias by combining the density and weak lensing fields}

\author[Arnau Pujol, Chihway Chang, Enrique Gazta\~{n}aga et al.]{\parbox{\textwidth}{Arnau Pujol\thanks{E-mail: pujol@ice.cat}$^{1}$, Chihway Chang$^{2}$, Enrique Gazta\~{n}aga$^{1}$, Adam Amara$^{2}$, \\
Alexandre Refregier$^{2}$, David J. Bacon${^3}$, Jorge Carretero$^{1,4}$, Francisco J. Castander$^{1}$, Martin Crocce$^{1}$, Pablo Fosalba$^{1}$, Marc Manera$^{5}$, Vinu Vikram$^{6,7}$}\vspace{0.4cm}\\
$^{1}$Institut de Ci\`encies de l'Espai (ICE, IEEC/CSIC), E-08193 Bellaterra (Barcelona), Spain\\
$^{2}$Department of Physics, ETH Zurich, Wolfgang-Pauli-Strasse 16, CH-8093 Zurich, Switzerland\\
$^{3}$Institute of Cosmology and Gravitation, University of Portsmouth, Dennis Sciama Building, Portsmouth, PO1 3FX, U.K.\\
$^{4}$Institut de F\'isica d'Altes Energies, Universitat Aut\`onoma de Barcelona, E-08193 Bellaterra, Barcelona, Spain\\
$^{5}$University College London, Gower Street, London, WC1E 6BT, U.K\\
$^{6}$Argonne National Laboratory, 9700 South Cass Avenue, Lemont, IL 60439, USA\\
$^{7}$Department of Physics and Astronomy, University of Pennsylvania, Philadelphia, PA 19104, USA\\
}
\date{Accepted xxxx. Received xxx}

\pagerange{\pageref{firstpage}--\pageref{lastpage}} \pubyear{2012}

\maketitle

\label{firstpage}

\begin{abstract}
We present a new method to measure the redshift-dependent galaxy bias by combining information from the galaxy density field and the weak lensing field. This method is based on \cite{Amara2012}, where they use the galaxy density field to construct a bias-weighted convergence field $\kappa_{g}$. The main difference between \cite{Amara2012} and our new implementation is that here we present another way to measure galaxy bias using tomography instead of bias parameterizations. The correlation between $\kappa_g$ and the true lensing field $\kappa$ allows us to measure galaxy bias using different zero-lag correlations, such as $\langle \kappa_g \kappa \rangle / \langle \kappa \kappa \rangle$ or $\langle \kappa_g \kappa_g \rangle / \langle \kappa_g \kappa \rangle$. Our method measures the linear bias factor on linear scales under the assumption of no stochasticity between galaxies and matter. We use the MICE simulation to measure the linear galaxy bias for a flux-limited sample ($i<22.5$) in tomographic redshift bins using this method. This paper is the first that studies the accuracy and systematic uncertainties associated with the implementation of the method, and the regime where it is consistent with the linear galaxy bias defined by projected 2-point correlation functions (2PCF). We find that our method is consistent with linear bias at the percent level for scales larger than 30 arcmin, while nonlinearities appear at smaller scales. This measurement is a good complement to other measurements of bias, since it does not depend strongly on $\sigma_{8}$ as the 2PCF measurements. We apply this method to the Dark Energy Survey Science Verification data in a follow-up paper.

\end{abstract}


\maketitle

\section{Introduction}

The formation and evolution of the large scale structures in the Universe is an important tool for cosmology studies. But since most of the mass in the Universe is in the form of dark matter, which cannot be directly observed, we need to understand the connection between the observable universe (galaxies and stars) and dark matter. In the $\Lambda$CDM paradigm, structures form in the initial density peaks  causing dark matter to gravitationally collapse  and form virialized objects. Galaxies are expected to follow these gravitational potentials \citep[e.g.][]{White1978}, and because of this they are tracers of the dark matter density peaks. The relation between the galaxy and mass distributions can be described theoretically with the galaxy bias prescription \citep{Kaiser1984,Fry1993,Bernardeau1996,Mo1996,Sheth1999,Manera2010,Manera2011}. Galaxy bias allows us to connect the distribution of galaxies with that of dark matter, and a good knowledge of galaxy bias would be very important to improve the precision of our cosmological measurements \citep{Eriksen2015}. 

Many papers have studied halo and galaxy bias in simulations \citep{Cole1989,Kravtsov1999,Seljak2004,Angulo2008,Faltenbacher2010,Tinker2010,Manera2011,Paranjape2013,Pujol2014,Zentner2014,Carretero2015,Pujol2015}, and the different ways to measure bias \citep{Kravtsov1999,Bernardeau2002,Manera2011,Roth2011,Pollack2014,Hoffmann2015,Bel2015}. There are also several measurements of bias in observations where usually the dark matter clustering is assumed from a model or from simulations \citep{Zehavi2011,Coupon2012,Cacciato2012,Jullo2012,Marin2013,Durkalec2014,DiPorto2014,Crocce2015b}. In most of these studies, however, the results depend strongly on assumptions of the cosmological parameters.

Gravitational lensing is the effect of light deflection due to the perturbations in the gravitational potential from mass distribution. It is a powerful tool to measure the mass distribution in the Universe, since the gravitational potential is affected by both baryonic and dark matter. Weak lensing refers to the statistical study of small distortions (around $1\%$) in the shapes of a large number of galaxies due to this effect. Several recent, ongoing and future galaxy surveys aim to obtain large weak lensing data sets that will allow us to better constrain cosmology, including the Canada-France-Hawaii Telescope Lensing Survey (CFHTLenS; \citealt{Heymans2012,Erben2013}), the Hyper Suprime-Cam (HSC; \citealt{Miyazaki2006}), the Dark Energy Survey (DES; \citealt{DES2005,Flaugher2005}), the Kilo Degree Survey (KIDS; \citealt{Kuijken2015}), the Panoramic Survey Telescope and Rapid Response System (PanSTARRS; \citealt{Kaiser2010}), the Large Synoptic Survey Telescope (LSST; \citealt{LSST2009}), Euclid \citep{Laureijs2011} the The Red Cluster Sequence Lensing Survey (RCSLenS; \citealt{Hildebrandt2016}), and Wide-Field Infrared Survey Telescope (WFIRST; \citealt{Green2012}). From the shape of the galaxies one can statistically infer the lensing fields, which contain information of the projected matter distribution and can be used to generate 2D and 3D mass maps \citep{Massey2007,VanWaerbeke2013, Vikram2015}. 

The combination of weak lensing and galaxy density information gives us a powerful handle for measuring galaxy bias. One way is by studying the cross-correlation between the aperture mass and number counts statistics, which are measurements of both dark matter and galaxy densities \citep{vanWaerbeke1998,Schneider1998}. In \cite{Hoekstra2002} they use the Red-Sequence Cluster Survey (RCS) and the VIRMOS-DESCART survey to measure galaxy bias at $z \simeq 0.35$ from the zero lag cross-correlation between aperture mass and number counts. They also find a scale dependence of bias on scales below $100$ arcmin. The same method has then been applied in more recent studies \citep{Simon2007,Jullo2012,Mandelbaum2013,Buddendiek2016}. Using a shear tomography analysis, \cite{Simon2012} combined galaxy-galaxy lensing and galaxy clustering to constrain the 3D galaxy biasing parameters. Bias can also be obtained from the cross-correlation between lensing from the Cosmic Microwave Background and the galaxy densities \citep{Giannantonio2016}. Using another method, \cite{Amara2012} (hereafter A12) used the COSMOS field to measure galaxy bias by reconstructing a bias-weighted shear map from the galaxy density field. Galaxy bias is estimated from the zero-lag cross correlation between this bias-weighted shear map from the galaxy density field and the shear measured from galaxy shapes. Different parameterizations of bias are used to measure constant, non-linear and redshift-dependent bias.  

In this paper we explore and extend the method from A12. We analyze whether the galaxy bias measured with our method is consistent with the linear bias obtained from the projected 2-point correlation functions (2PCF). We find that our method can be affected by different parameters in the implementation such as redshift binning, the redshift range used, angular scales, survey area and shot noise.  Finally, we show how to measure the redshift-dependent galaxy bias by using tomographic redshift binning. Although this method is very similar to the one presented in A12, there are few notable differences. First of all, in A12 they explore different smoothing schemes for the density field, while we explore pixelizing the maps and applying a Top Hat filter. In A12 the lensing shear is estimated for each galaxy, and the bias is measured from the predicted and measured shear of the galaxies, while we measure galaxy bias from the generated lensing maps. Finally, A12 fit different parametric biases using a wide range of redshift for the galaxy density field, while here we implement a tomographic measurement, where we measure bias in redshift bins by using the density field of galaxies in each particular bin. 
We apply this method to the DES Science Verification (SV) data in a second paper (\citealt{Chang2016}, hereafter Paper II). 

The paper is organized as follows. In \S\ref{sec:background} we give an overview of the theory for our analysis. In \S\ref{sec:method} we present the method used to measure bias from the galaxy density and weak lensing fields and the numerical effects associated with the implementation of the method. In \S\ref{sec:results} we present the results of the different tests and the final measurement of redshift-dependent galaxy bias. We finally close in \S\ref{sec:conclusions} with discussion and conclusions.

\section{Theory}\label{sec:background}

\subsection{Galaxy Bias}

The distribution of galaxies traces that of dark matter, and one of the common descriptions for this relation is galaxy bias, which relates the distribution of galaxies with that of dark matter. There are several ways to quantify galaxy bias \citep{Bernardeau2002,Manera2011,Roth2011,Hoffmann2015,Bel2015}, and one of the most common ones is from the ratio of the 2PCFs of galaxies and dark matter: 
\begin{equation}
\xi_g(r) = b^2(r) \xi(r), 
\label{eq:bias_2pcf_3d}
\end{equation}
where $b(r)$ is the galaxy bias, and $\xi_g(r)$ and $\xi(r)$ are the scale-dependent galaxy and matter 2PCFs respectively, which are defined as:

\begin{equation}
\xi_g(r_{12}) = \langle \delta_g(\bm{r_1}) \delta_g(\bm{r_2}) \rangle,\  \xi(r_{12}) =  \langle \delta(\bm{r_1}) \delta(\bm{r_2}) \rangle.
\end{equation}
where $\delta_g=(\rho_{g} - \bar{\rho}_{g})/\bar{\rho}_{g}$ is the density fluctuation of galaxies ($\rho_{g}$ is the galaxy number density), and  $\delta=(\rho - \bar{\rho})/\bar{\rho}$ is the density fluctuation of dark matter ($\rho$ is the dark matter density). As can be seen from this equation, galaxy bias generally depends on the scale $r_{12}$ (defined as the distance between $\bm{r_1}$ and $\bm{r_2}$). However, it has been shown that at sufficiently large scales in the linear bias regime, bias is constant \citep[e.g.][]{Manera2011}. 

Bias can also be defined from the projected 2PCFs: 
\begin{equation}
\omega_g(\theta) = b^2(\theta) \omega(\theta), 
\label{eq:bias_2pcf}
\end{equation}
where $\omega_g(\theta)$ and $\omega (\theta)$ refer to the projected 2PCF of galaxies and dark matter respectively. This definition of bias will be used in the analysis of this paper. In this case, the bias dependence is on separation angle $\theta$ instead of distance $r$.

In the local bias model approach \citep{Fry1993}, the density field of galaxies is described as a function of its local dark matter density, so that $\delta_g = F[\delta]$. We can express this relation as a Taylor series:

\begin{equation}
\delta_g = \epsilon + b_0 + b_1 \delta + \frac{b_2}{2} \delta^2 + ... = \sum_{i=0}^{\infty} b_i(z) \delta^i + \epsilon,
\end{equation}
where $b_{i}$ are the coefficients of the Taylor expansion and $\epsilon$ represents the galaxy shot noise. The density contrasts $\delta_g$ and $\delta$ are smoothed to a certain scale by a window function, so the relation also depends on that physical scale. It also assumes no random scatter between $\delta_g$ and $\delta$, and $\epsilon$ is negligible for large smoothing scales. In the linear regime, $\delta \ll 1$, and as $b_0 = 0$ because $\langle \delta_g \rangle = \langle \delta \rangle = 0$, then the equation becomes: 
\begin{equation}
\delta_g = b_1 \delta
\label{eq:bias_delta}
\end{equation}

According to \cite{Manera2011}, at large scales this definition of bias is consistent with the bias obtained from the 2PCFs: for $r_{12} \gtrsim 40 \Mpc$, $b$ from equation (\ref{eq:bias_2pcf_3d}) is approximately constant and consistent with $b_1$ from equation (\ref{eq:bias_delta}). This $b_1$ can then be measured from the different zero-lag correlations between $\delta_g$ and $\delta$:
\begin{equation}
b_1 = \frac{\langle \delta_g \delta \rangle}{\langle \delta \delta \rangle} 
\label{eq:local_bias_corrs1}
\end{equation}
\begin{equation}
b_1 = \frac{\langle \delta_g \delta_g \rangle}{\langle \delta_g \delta \rangle} 
\label{eq:local_bias_corrs2}
\end{equation}
\begin{equation}
b_1  = \sqrt{\frac{\langle \delta_g \delta_g \rangle}{\langle \delta \delta \rangle}}
\label{eq:local_bias_corrs3}
\end{equation}

Although these relations appear to measure the same parameter $b_1$, the results can be affected by the stochasticity in the relation between $\delta_g$ and $\delta$, that can come from different effects, such as the stochasticity of bias and the projection effects. 

Galaxy bias from equations (\ref{eq:local_bias_corrs1}-\ref{eq:local_bias_corrs3}) depend on the smoothing scale used to measure $\delta$ and $\delta_g$. For small scales nonlinearities in the relation between $\delta$ and $\delta_g$ appear, and $b_1$ is no longer consistent with equation (\ref{eq:bias_2pcf}). Throughout the paper we will use these equations of bias for distributions projected in the sky. Then, the relations in this analysis depend on angular distance (for equation (\ref{eq:bias_2pcf})) or smoothing angle (for equations (\ref{eq:local_bias_corrs1}-\ref{eq:local_bias_corrs3})). The relation between both scales of bias (smoothing and separation) is complex, since the smoothing of $\delta$ and $\delta_g$ on a scale $\Theta$ involves the correlations of all the scales below $\Theta$. However, in the linear and local regime bias is consistent with both scales and then all the estimators can be compared.

\subsection{Weak Lensing}

Weak gravitational lensing \citep[see e.g.][]{Bartelmann2001,Refregier2003} measures the small changes of galaxy shapes and brightnesses due to the foreground mass distribution in the line-of-sight of the (source) galaxies. By studying this effect statistically, assuming that (lensed) galaxies are randomly oriented in the absence of lensing, one can infer the mass distribution in the foreground of these source galaxies. As the light distortion is affected by gravity, weak lensing allows us to measure the total mass distribution, including baryonic and dark matter. 

The gravitational potential $\Phi$ of a given density distribution $\delta$ can be defined as: 
\begin{equation}
\nabla^2 \Phi = \frac{3 H_0^2 \Omega_m}{2a} \delta,
\label{eq:grav_pot}
\end{equation}
where $H_0$ and $\Omega_m$ are the Hubble parameter and the matter density parameter today respectively, and $a$ is the scale factor assuming a spatially flat Universe. Assuming General Relativity and no anisotropic stress, the lensing potential for a given source at position $(\bm\theta, \chi_s)$ is given by the weighted line-of-sight projection of $\Phi$:
\begin{equation}
\psi\left(\bm\theta, \chi_s \right) = 2 \int_0^{\chi_s}{\mathrm{d}\chi \frac{\chi (\chi_s - \chi)}{\chi_s} \Phi \left(\bm\theta, \chi \right)},
\label{eq:lens_pot}
\end{equation}
where $\bm\theta$ is the angular position on the sky, $\chi$ refers to the comoving radius and $\chi_s$ is the comoving distance to the sources. The distortion of the source galaxy images can be described by the convergence $\kappa$ and shear $\bm\gamma$ fields that are defined as: 
\begin{equation}
\kappa = \frac{1}{2}\nabla^2 \psi,
\label{eq:kappa}
\end{equation}
\begin{equation}
\bm{\bm\gamma} = \gamma_1 + i \gamma_2 = \frac{1}{2}(\psi_{,11} - \psi_{,22}) + i \psi_{,12},
\label{eq:gamma}
\end{equation}
where $\psi_{,ij} = \partial_i \partial_j \psi$. Focusing on the convergence field, combining equations (\ref{eq:grav_pot}), (\ref{eq:lens_pot}) and (\ref{eq:kappa}) we obtain: 
\begin{equation}
\kappa(\bm{\theta}, \chi_s) = \frac{3H_0^2 \Omega_m}{2c^2} \int_0^{\chi_s}{d \chi \frac{\chi (\chi_s - \chi)}{\chi_s} \frac{\delta(\bm{\theta}, \chi)}{a(\chi)}} \equiv K [\delta]
\label{eq:kappa_int}
\end{equation}

For simplicity, we define $q (\chi, \chi_s)$ as the lensing kernel of the integral of $\delta$ at $\chi$ with the source at $\chi_s$:
\begin{equation}
q(\chi, \chi_s) = \frac{3H_0^2 \Omega_m}{2c^2} \frac{\chi (\chi_s - \chi)}{\chi_s a(\chi)}
\label{eq:q_comoving}
\end{equation}
so that
\begin{equation}
\kappa(\bm{\theta}, \chi) = \int_0^{\chi_s}{ q(\chi, \chi_s) \delta(\bm\theta, \chi)d\chi}.
\label{eq:simple_kappa_int}
\end{equation}
Note that $\kappa$ corresponds to a weighted integral of the matter density fluctuations in the line-of-sight of the source galaxies.

\section{method}\label{sec:method}

\subsection{Simulation}

For the analysis we use the MICE Grand Challenge simulation \citep{Fosalba2015,Fosalba2015b,Crocce2015}, an N-body simulation of a $\Lambda$CDM cosmology with the following cosmological parameters: $\Omega_m = 0.25$, $\sigma_8 = 0.8$, $n_s = 0.95$, $\Omega_b = 0.044$, $\Omega_\Lambda = 0.75$, $h = 0.7$. It has a volume of $(3.072\Gpc)^3$ with $4096^3$ particles of mass $2.927\times 10^{10}\Mo$. The galaxy catalogue has been run according to a Halo Occupation Distribution (HOD) and a SubHalo Abundance Matching (SHAM) prescriptions \citep{Carretero2015}. The parameters of the model have been fitted to reproduce clustering as a function of luminosity and colour from the Sloan Digital Sky Survey \citep{Zehavi2011}, as well as the luminosity function \citep{Blanton2003,Blanton2005b} and colour-magnitude diagrams \citep{Blanton2005c}. We use the MICECATv2 catalogue, an extension of the publicly available MICECATv1 catalogue\footnote{\texttt{http://cosmohub.pic.es/}}.  The main difference between MICECATv1 and MICECATv2 is that MICECATv2 is complete for $i<24$ from $z = 0.07$ to $z = 1.4$, while MICECATv1 is complete for an absolute magnitude of $M_r < -19$. The catalogue also contains the lensing quantities ($\gamma_1$, $\gamma_2$ and $\kappa$) at the position of each galaxy, calculated from the dark matter field with a resolution of Nside=8192 in healpix (corresponding to a pixel size of $\sim 0.43$ arcmin). The lensing signal was computed using the Born approximation. As the lensing value assigned to a galaxy at a given 3D position is inherited from the corresponding pixel value of the dark matter lensing map in which that galaxy sits in, the lensing quantities of the galaxies do not have shape noise. 

\subsection{Bias estimation}\label{sec:bias_method}

In this section, we introduce the method used to estimate galaxy bias from the lensing and density maps of galaxies in the MICE simulation. It consists on the construction of a template $\kappa_{g}$ for the lensing map $\kappa$ from the density distribution of the foreground galaxies assuming equation (\ref{eq:bias_delta}). Substituting $\delta$ with $\delta_g$ in equation (\ref{eq:kappa_int}) gives:
\begin{equation}
\kappa_g (\bm\theta) = \int_0^{\chi_s}{ q(\chi, \chi_s) \delta_g(\bm\theta, \chi)d\chi}
\label{eq:kg}
\end{equation}
When computing $\kappa_g$ numerically, the integral is approximated by a sum over all lenses in the foreground of the sources: 
\begin{equation}
\kappa_g (\bm\theta) \simeq \sum_{i = 1}^N{ \bar{q}^i \delta^i_g(\bm\theta)\Delta \chi^i},
\label{eq:kg_sum}
\end{equation}
where we have split the foreground galaxies into $N$ redshift bins. $\Delta \chi^i$ refers to the redshift bin width of the $i$th bin in comoving coordinates, $\bar{q}^i$ is the mean lensing weight that corresponds to that redshift bin and $\delta^i_{g}(\bm\theta)$ is the galaxy density fluctuation in that redshift bin at position $\bm\theta$, where $\bm\theta$ now represents a pixel in the sky plane. $\delta^i_{g}(\bm\theta)$ is calculated through $\delta^i_{g} (\bm\theta) = (\rho^i_{g} (\bm\theta) -\bar{\rho}^i_{g})/\bar{\rho}^i_{g}$, where $\rho^i_{g} (\bm\theta)$ is the density of galaxies projected in the line-of-sight in the $i$th redshift bin and position (pixel) $\bm\theta$, and $\bar{\rho}^i_{g}$ is the mean density of galaxies in the redshift bin, calculated from all the galaxies inside the redshift bin. This measurement of $\bar{\rho}^i_{g}$ gives a good estimate of the mean density if the redshift bin is wide enough. For narrow bins of redshift width below $\Delta z = 0.03$ a smoothing of $\bar{\rho}^i_{g}$ as a function of redshift is needed to obtain a good estimate of the mean density, as discussed in \S\ref{sec:implementation}. Notice that $\delta^i_{g}(\bm\theta)$ is calculated taking into account all the galaxies inside the volume of the cell corresponding to each pixel and redshift bin. This means that it corresponds to a projection in redshift of the galaxy density field weighted by the volume of the corresponding cell.

\begin{figure}
\centering
\includegraphics[scale=.58]{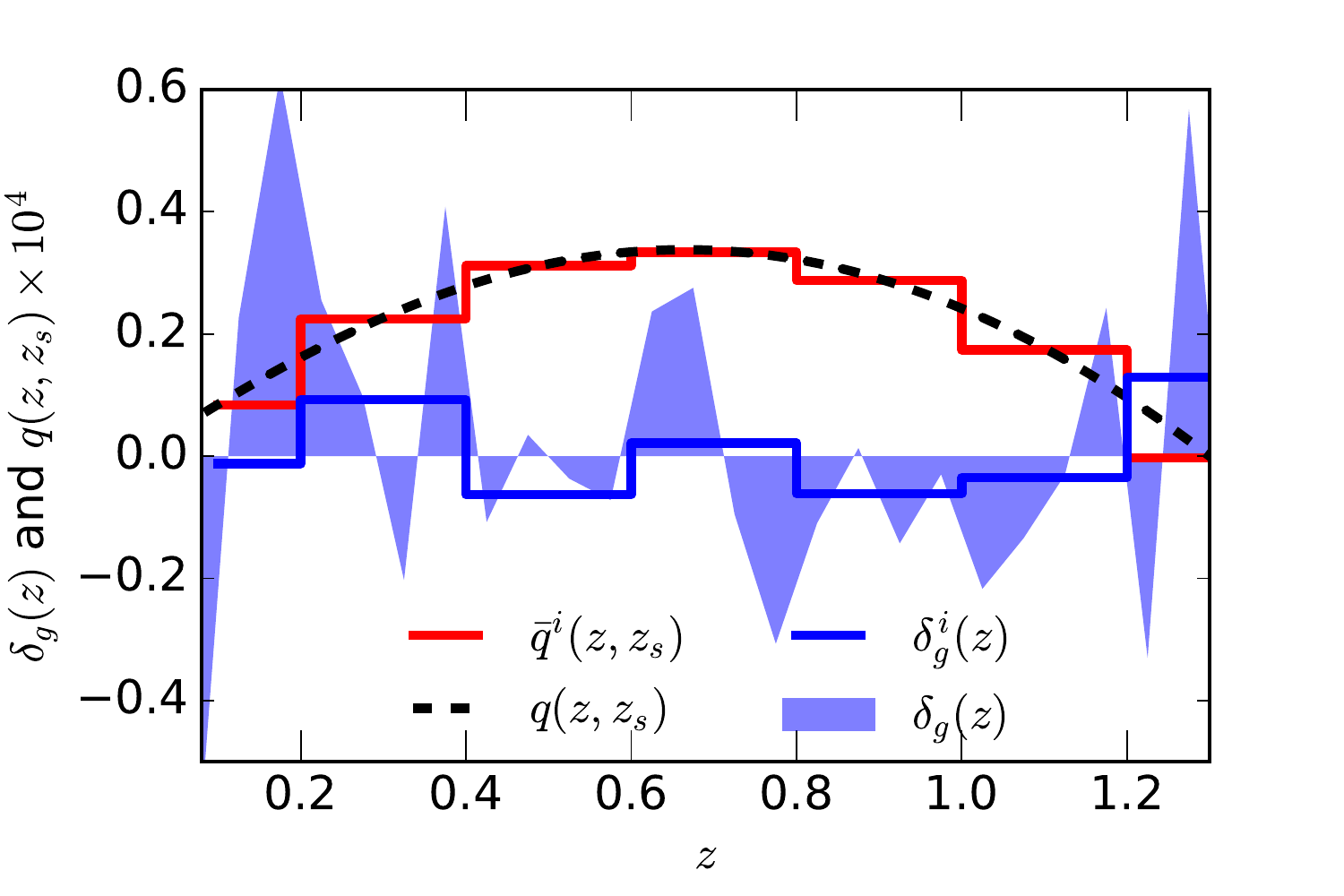}
\caption{Schematic comparison of equations (\ref{eq:kg},\ref{eq:kg_sum}). Dahsed black line shows $q(z,z_s)$, defined in comoving scales in equation (\ref{eq:q_comoving}), for a fixed $z_s = 1.3$, while red solid line shows $\bar{q}^i$ from equation (\ref{eq:kg_sum}) in redshift bins of $\Delta z = 0.2$. The blue shaded region represents $\delta_g(z)$ using narrow redshift bins ($\Delta z = 0.05$). The blue solid line represents $\delta^i_g$ for the redshift bins of $\Delta z = 0.2$.}\label{fig:delta_vs_z}
\end{figure}

In Figure \ref{fig:delta_vs_z} we show a schematic picture of the effects of equation (\ref{eq:kg_sum}). Dashed black line shows $q(z,z_s)$, defined from equation (\ref{eq:q_comoving}) in redshift coordinates, while red solid line shows $\bar{q}^i$ in redshift bins of $\Delta z = 0.2$. We used $z_s = 1.3$ for this figure. The blue shaded region represents $\delta_g(z)$ in a random (just for the example) pixel in the sky using narrow redshift bins ($\Delta z = 0.05$). The blue solid line represents  $\delta^i_g$ for the redshift bins of $\Delta z = 0.2$. Equation (\ref{eq:kg_sum}) then is equivalent to the integral of the product of the blue and red solid lines. 

Equation (\ref{eq:kg_sum}) is an approximation of (\ref{eq:kg}), that assumes that the small fluctuations in redshift of $\delta_g$ inside the bins do not affect the results. The mean of $q(\chi, \chi_s) \delta_g(\chi)$ inside the bins can be approximated by the product of the means $\bar{q}^i \delta^i_g(\bm\theta)$. These approximations hold at large scales and when $q(\chi, \chi_s)$ and $\delta_g(\chi)$ are not correlated. 

We focus on the simplest case, where galaxy bias is linear, local and redshift-independent. In this case, we can estimate $b$ from the following zero-lag correlations of $\kappa$ and $\kappa_g$: 
\begin{equation}
b = \frac{\langle \kappa_g \kappa \rangle}{\langle \kappa \kappa \rangle - \langle \kappa^N \kappa^N \rangle} 
\label{eq:zero_lag1}
\end{equation}
\begin{equation}
b = \frac{\langle \kappa_g \kappa_g \rangle - \langle \kappa_g^N \kappa_g^N \rangle}{\langle \kappa_g \kappa \rangle},
\label{eq:zero_lag2}
\end{equation}
where $\kappa^N$ and $\kappa^N_g$ are the sampling and shot-noise correction factors obtained by randomizing the galaxy positions and re-calculating $\kappa$ and $\kappa_g$. $\kappa$ is obtained from the mean $\kappa$ of the galaxies in each pixel. This is affected by the number of source galaxies in the pixel, causing a noise in $\langle \kappa \kappa \rangle$ that depends on the angular resolution used, reaching a $10\%$ error for a pixel size of 5 arcmin. This noise is cancelled by subtracting $\langle \kappa^N \kappa^N \rangle$. On the other hand, $\langle \kappa_g \kappa_g \rangle$ is affected by shot noise, causing an error that increases with the angular resolution up to a $20\%$ for a pixel size of 5 arcmin. This noise is cancelled by subtracting $\langle \kappa_g^N \kappa_g^N \rangle$. This correction assumes a Poisson distribution. To test how well this correction works for this method, we calculated $\langle \kappa_g \kappa_g \rangle - \langle \kappa_g^N \kappa_g^N \rangle$ using the dark matter particles instead of galaxies, and we compared the results with the true $\langle \kappa \kappa \rangle$ maps from the simulation. We did this with different dilutions (from $1/70$ to $1/700$) of the dark matter particles, and recover $\langle \kappa \kappa \rangle$ better than $1\%$ independently on the dilution, indicating that the shot-noise subtraction is appropriate.

Since the galaxies used from the MICE simulation do not have shape noise, the estimators in this analysis are not affected by shape noise. This is not the case in observations, where shape noise is the most important source of noise of this method and needs to be corrected. Moreover, in observations we do not have $\kappa$ either, and we need to obtain $\kappa$ from $\bm{\gamma}$ and equations (\ref{eq:kappa}-\ref{eq:gamma}) in order to use these estimators. Notice that galaxy bias obtained from equations (\ref{eq:zero_lag1}-\ref{eq:zero_lag2}) imply an average of bias as a function of redshift. This is because $\kappa_g$ involves a redshift integral of $\delta_g \sim b(z) \delta$ as specified in equation (\ref{eq:kg}), so the final product is a redshift-averaged bias weighted by the lensing kernels that appear in equations  (\ref{eq:zero_lag1}-\ref{eq:zero_lag2}). Later in this analysis we use tomographic redshift bins, where we assume that bias does not significantly change inside the bin, and we measure bias in each of the redshift bins.

\begin{figure}
\centering
\includegraphics[scale=.43]{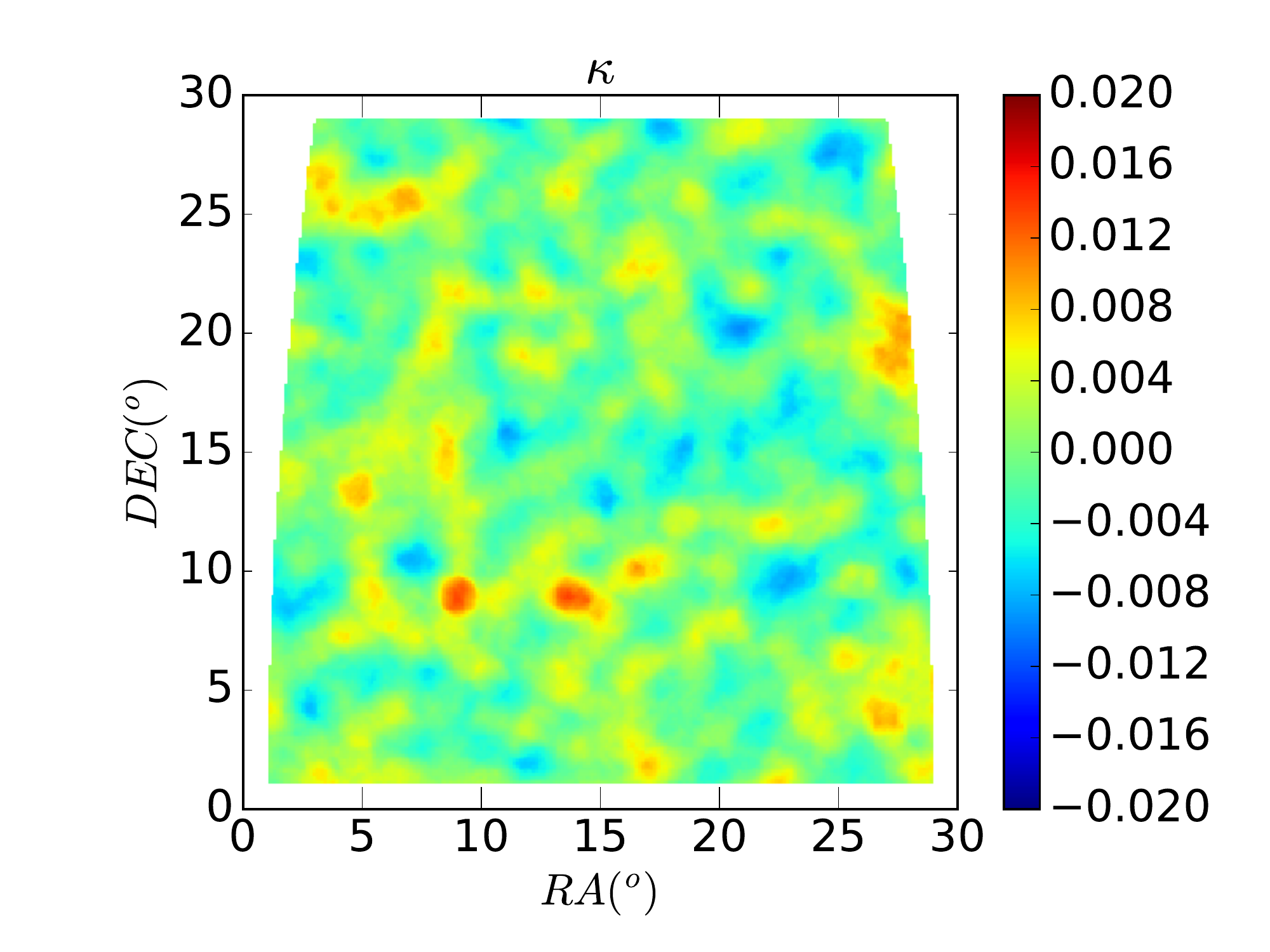}
\includegraphics[scale=.43]{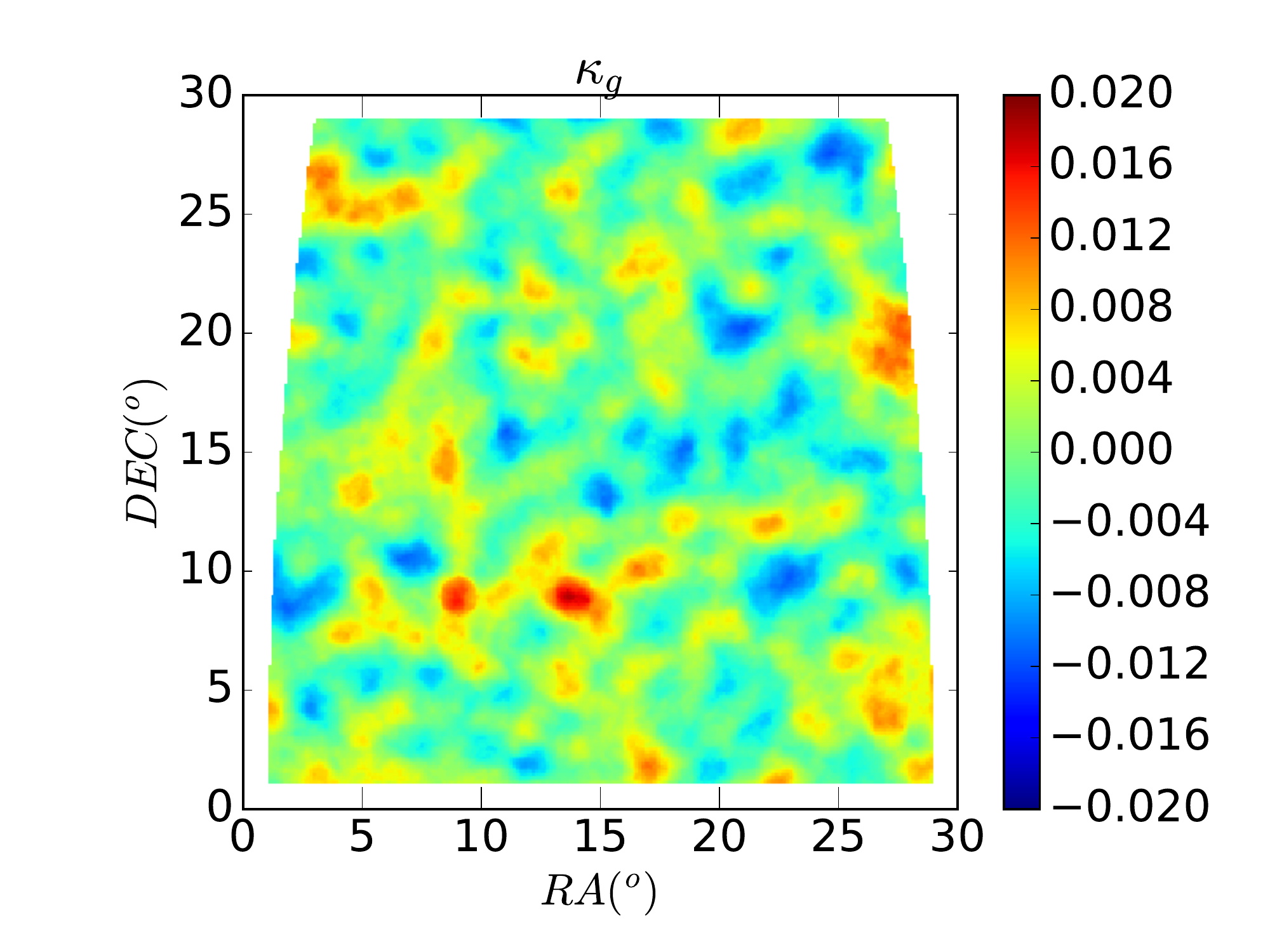}
\includegraphics[scale=0.43]{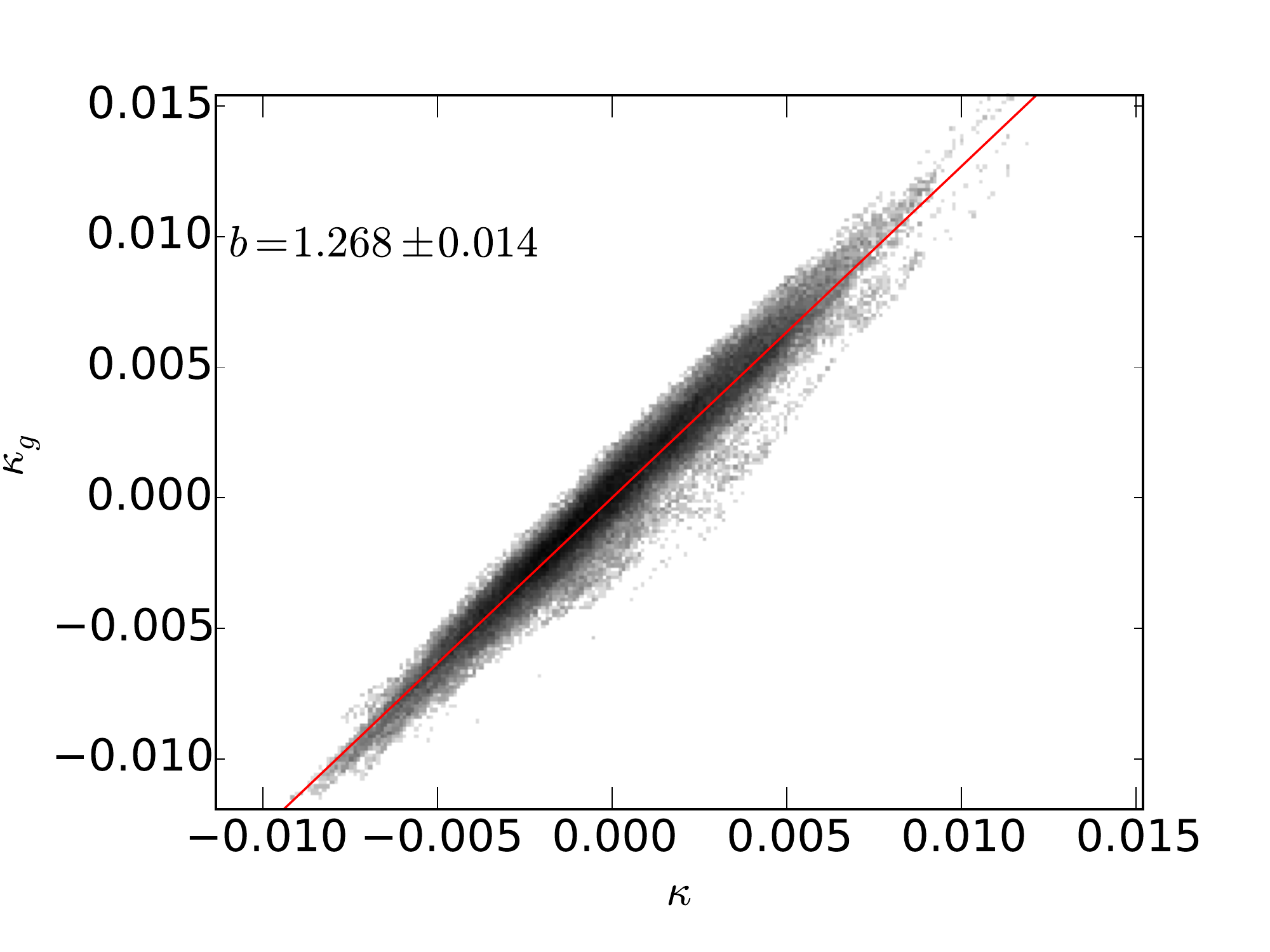}
\caption{Comparison of $\kappa$ vs $\kappa_g$. Top panel shows the $\kappa$ field from the source galaxies within $0.9<z<1.1$ and using a Top Hat filter of $50$ arcmin of radius. Middle panel shows $\kappa_g$ obtained from equation (\ref{eq:kg_sum}), using the same smoothing scheme. Bottom panel shows the comparison between $\kappa_g$ and $\kappa$ for the pixels of the maps, with the specified bias and error obtained. The red line corresponds to a line crossing the origin and its slope corresponds to $b$. It is consistent with the linear fit of the distribution of the points.}\label{fig:ideal_case_constant}
\end{figure}

To measure the errors on $b$, we use the Jackknife (JK) method. We divide the area into $16$ subsamples. We evaluate $b$ $16$ times excluding each time a different subsample. The error of $b$ is estimated from the standard deviation of these $16$ measurements as:
\begin{equation}
\sigma(b) \simeq \sqrt{ \frac{N_{JK} - 1}{N_{JK}}  \sum^{N_{JK}}_{i = 1} (b_i - b)^2 },
\label{eq:jk_error}
\end{equation}
where $N_{JK}$ refers to the number of JK subsamples used, $b_i$ is the bias measured by excluding the $ith$ subsample and $b$ is obtained from the average overall subsamples. We checked that the error are very similar if we use a different number of subsamples (between $9$ and $100$) instead of $16$.

Note that we can also measure bias from the following cross correlations: 
\begin{equation}
b = \frac{\langle \gamma_{i,g} \gamma_i \rangle}{\langle \gamma_i \gamma_i \rangle - \langle \gamma_i^N \gamma_i^N \rangle} 
\label{eq:zero_lag_gamma1}
\end{equation}
\begin{equation}
b =  \frac{\langle \gamma_{i,g}\gamma_{i,g}\rangle - \langle \gamma_{i,g}^N \gamma_{i,g}^N \rangle}{\langle \gamma_{i,g}^N\gamma_{i}^N\rangle}, i = 1,2
\label{eq:zero_lag_gamma2}
\end{equation}
As this is not the focus of the paper, and we can obtain $\kappa$ from the simulation, we measure $b$ from equations (\ref{eq:zero_lag1},\ref{eq:zero_lag2}) in this study. However, in observations we measure the shape of the galaxies, that is directly related to $\gamma_i$. Because of this, applying this method to data requires a conversion from $\kappa_g$ to $\gamma_{i,g}$ or from $\gamma_i$ to $\kappa$. These conversions imply other systematics due to the finite area and the irregularities of the mask. The conversion from $\kappa_g$ to $\gamma_{i,g}$ can be affected by the shot noise in $\kappa_g$, but this noise is less dominant than shape noise, that can affect the conversion from $\gamma_i$ to $\kappa$. We address this issue in Paper II, where we use conversions based on \cite{Kaiser1993} (hereafter KS method) to apply this method to DES SV data. Another aspect to take into account for data analysis is that since shape noise is the main source of noise in the measurement, we like to avoid the terms that involve variance
of lensing quantities  $\langle \kappa \kappa \rangle$ and $\langle \gamma_i\gamma_i\rangle$, since these terms are the most affected by shape noise.

\subsection{Implementation}\label{sec:implementation}

In Figure \ref{fig:ideal_case_constant}, we illustrate our procedure. We used a $\sim900$ square degree area from the MICE simulation corresponding to $0^{\circ}<RA<30^{\circ}$ and $0^{\circ}<DEC<30^{\circ}$. The top panel shows the convergence map $\kappa$ of source galaxies located at $z \simeq 1$. The middle panel shows the constructed convergence template, $\kappa_g$, derived via equation (\ref{eq:kg_sum}). Both maps have been generated by pixelizing the distributions in pixels of $7$ arcmin of side. Then, the pixelated maps have been smoothed using a circular top hat filter of $50$ arcmin of radius from this pixelated map. The map obtained corresponds to the angular scale of $50$ arcmin, and their statistics do not depend on the scale of the previous pixelization (if the pixels are much smaller than the smoothing scale). We can see that $\kappa_g$ is a biased version of $\kappa$ at large scales. In the bottom panel we show the scatter plot of $\kappa$ versus $\kappa_g$. The bias $b$ shown in the plot is estimated via equation (\ref{eq:zero_lag1}), and the error corresponds to the JK errors from equation (\ref{eq:jk_error}). In red, we show a line crossing the origin and with the slope corresponding to this estimated bias. We have checked that the $b$ value derived from the zero-lag statistics is in agreement with a linear fit to the scatter plot at the $0.1\%$ level. This is an indication that we are in the linear regime, where we can assume equation (\ref{eq:bias_delta}).

We note that the expression for the bias from equations (\ref{eq:zero_lag1},\ref{eq:zero_lag2}) assumes equation (\ref{eq:bias_delta}). 
However, $\kappa$ is a projection of $\delta$ in the line-of-sight weighted by the lensing kernel, as well as $\kappa_g$. Thus, the relation between $\kappa$ and $\kappa_g$ is a constant that comes from the redshift dependence of bias weighted by the redshift dependence of the lensing kernel. Hence, the bias obtained in this example is a weighted mean of galaxy bias as a function of redshift. But we can take this dependence into account to measure bias at different redshifts using tomography as we explain in \S\ref{sec:z_dependence} below.

\subsection{Redshift dependence}\label{sec:z_dependence}

This method involves an integral (or a sum in practice) along the redshift direction, and because of this the bias obtained is a weighted average of the redshift dependent bias. However, we can estimate galaxy bias in a given redshift bin if we restrict the calculation to the foreground galaxies in that redshift bin, assuming that bias does not change significantly in the bin. If this is the case, we can measure the redshift-dependent bias using tomographic redshift bins. 

Since $\kappa_g$ is obtained from the contribution of all the galaxies in front of the sources, if we restrict the redshift range for the calculation of $\kappa_g$ we need to renormalize the result by taking into account the contribution from the unused redshift range.

 We define as $\kappa'_g$ the construction of a partial $\kappa_g$ using only the galaxies projected in a given redshift bin, so:
\begin{equation}
\kappa'_g(\bm\theta) = \int{ d\chi q(\chi, \chi_s)  p(\chi) \delta_g(\bm\theta, \chi)  },
\label{eq:partial_kg_exact}
\end{equation}
where $p(\chi)$ is the radial selection function, equal to 1 inside the bin $\chi_{min} < \chi < \chi_{max}$ and $0$ outside. To simplify the notation, when the limits are not specified in the integral, the integral will go through the whole range between $0$ and $\infty$. We assume that all the sources at located at $\chi_s$ here and in the following, and because of this we will not include the argument $\chi_s$ in $\kappa'_g(\theta)$ and other functions. Note that, as $p(\chi) = 0$ for all $\chi$ outside the bin, only the range $\chi_{min} < \chi < \chi_{max}$ contributes to the integral in equation (\ref{eq:partial_kg_exact}), and $p(\chi)$ implies a projection inside the bin. In order to simplify the expression, if $q(\chi, \chi_s)$ is not correlated with $p(\chi)\delta_g(\chi)$ inside the bin (which is the case, since $\delta_g(\chi)$ decorrelates quickly in the redshift direction and hence the correlation is only important for very narrow bins), $q(\chi, \chi_s)$ can be described outside the integral as: 
\begin{equation}
\kappa'_g(\bm\theta) \simeq  \bar{q}' \Delta\chi \int{ d\chi p'(\chi) \delta_g(\bm\theta, \chi) } = \bar{q}' \Delta\chi \bar{\delta}'_g, 
\label{eq:partial_kg}
\end{equation}
with
\begin{equation}
\bar{q}' = \int_{\chi_{min}}^{\chi_{max}}{ d\chi \frac{q(\chi, \chi_s)}{\Delta \chi} }.
\label{eq:mean_q}
\end{equation}
$\Delta \chi = \chi_{max} - \chi_{min}$, and now $p'(\chi)$ is the same selection function as $p(\chi)$ but normalized to 1, so $p'(\chi) = p(\chi)/\Delta \chi$. 

 With this definition, we measure the galaxy bias in this redshift bin, that we call $b'$, from the following expressions:
\begin{equation}
b_1' = \frac{1}{f_1} \frac{\langle \kappa'_g \kappa \rangle}{\langle \kappa \kappa \rangle - \langle \kappa^N \kappa^N \rangle}  
\label{eq:predicted_bias1}
\end{equation}
\begin{equation}
b_2' = \frac{1}{f_2} \frac{\langle \kappa'_g \kappa'_g \rangle - \langle {\kappa'_g}^{N} {\kappa'_g}^{N} \rangle}{\langle \kappa'_g \kappa \rangle},
\label{eq:predicted_bias2}
\end{equation}
where ${\kappa'_g}^{N}$ is obtained by randomizing the positions of the galaxies in the redshift bin in order to correct for shot-noise, and $f_1$ and $f_2$ correspond to the following ratios:
\begin{equation}
f_1 = \frac{\langle \kappa' \kappa \rangle}{\langle \kappa \kappa \rangle}
\label{eq:f_1}
\end{equation}
and 
\begin{equation}
f_2 = \frac{\langle \kappa' \kappa' \rangle}{\langle \kappa' \kappa \rangle},
\label{eq:f_2}
\end{equation}
where $\kappa'$ is defined as the contribution to $\kappa$ of the dark matter field projected in the redshift bin used, so: 
\begin{equation}
\kappa'(\bm\theta) = \bar{q}' \Delta\chi \int{ d\chi p'(\chi) \delta(\bm\theta, \chi) } = \bar{q}' \Delta\chi\bar{\delta}' .
\label{eq:partial_k}
\end{equation}

For our purpose we are interested in the analytic expressions of $\langle \kappa' \kappa \rangle$, $\langle \kappa' \kappa' \rangle$ and $\langle \kappa \kappa\rangle$ to be able to use $f_1$ and $f_2$ to measure galaxy bias in tomographic redshift bins. 
According to the definitions, from equations (\ref{eq:simple_kappa_int},\ref{eq:partial_k}) we can derive: 
\begin{equation}
\langle \kappa' \kappa (\theta) \rangle = \bar{q}' \Delta \chi' \int{ p'(\chi_1) d\chi_1} \int_0^{\chi_s}{ d \chi_2 q(\chi_2) \xi(r_{12})}
\label{eq:kprime_k}
\end{equation}
\begin{equation}
\langle \kappa' \kappa' (\theta) \rangle = (\bar{q}' \Delta \chi' )^2 \int{ p'(\chi_1) d\chi_1} \int{ d\chi_2 p'(\chi_2) \xi(r_{12})} 
\label{eq:kprime_kprime}
\end{equation}
\begin{equation}
\langle \kappa \kappa (\theta) \rangle = \int_{0}^{\chi_s}{ q(\chi_1) d\chi_1}\int_0^{\chi_s}{q(\chi_2) d\chi_2}\xi(r_{12}),
\label{eq:k_k}
\end{equation}
with $r_{12}^2 = \chi_1^2 + \chi_2^2 + 2 \chi_1 \chi_2 \cos \theta$, $\xi(r_{12})$ is the 2PCF and $\theta$ is the angular separation between the two fields.

For the general case, the zero-lag correlation of two fields $A$ and $B$ at an angular scale $\Theta$ (corresponding to a radius $R$ in the given redshift bin) is given by:
\begin{equation}
\langle \kappa_A \kappa_B (\Theta) \rangle = \frac{4 }{\pi R^{4}} \int_{0}^{R} dr_{1} r_{1}  \int_{0}^{R} dr_{2} r_{2}
\int_{0}^{\pi} d\eta \omega_{AB} (\theta),
\end{equation}
where $\theta^{2}=r_{1}^{2}+r_{2}^{2}-2r_{1}r_{2}\cos \eta$, $\kappa_{A}$ and $\kappa_{B}$ can be $\kappa$, $\kappa'$, $\kappa_{g}$ or $\kappa'_{g}$, $\eta$ is the angular separation between the vectors $\bm{r_1}$ and $\bm{r_2}$ and $\omega (\theta)$ is a projected two-point angular 
correlation function of the two fields $A$ and $B$ defined as:
\begin{equation}
\omega_{AB}(\theta) = \int d\chi_{A}\int d\chi_{B} q(\chi_A) q(\chi_B) p(\chi_A) p(\chi_B) \xi_{AB}(r),
\end{equation}
where $p(\chi_{A,B})$ are the corresponding selection functions of the fields $A$ and $B$, and $\xi_{AB}(r)$ is the 3D two-point cross-correlation function, that in this case corresponds to the dark matter $\xi(r)$.

In order to be consistent with equations (\ref{eq:partial_kg},\ref{eq:partial_k}), when $A$ (and also $B$) refer to the dark matter field limited in a redshift bin, we use the following expressions for the angular correlation functions:
\begin{equation}
\omega_{A' B}(\theta) = \bar{q}' \Delta \chi \int d\chi_{A}\int d\chi_{B}  q(\chi_B) p'(\chi_A) p(\chi_B) \xi_{AB}(r)
\end{equation}
\begin{equation}
\omega_{A' B'}(\theta) = \bar{q}'^2 \Delta \chi^2 \int d\chi_{A}\int d\chi_{B}  p'(\chi_A) p'(\chi_B) \xi_{AB}(r),
\end{equation}
where $A'$ and $B'$ refer to the cases where the fields $A$ and $B$ are restricted to the redshift bin, and $\Delta \chi = \chi_{max} - \chi_{min}$ defines the redshift bin width of $A'$ and $B'$. 

\begin{figure}
\centering
\includegraphics[scale=.58]{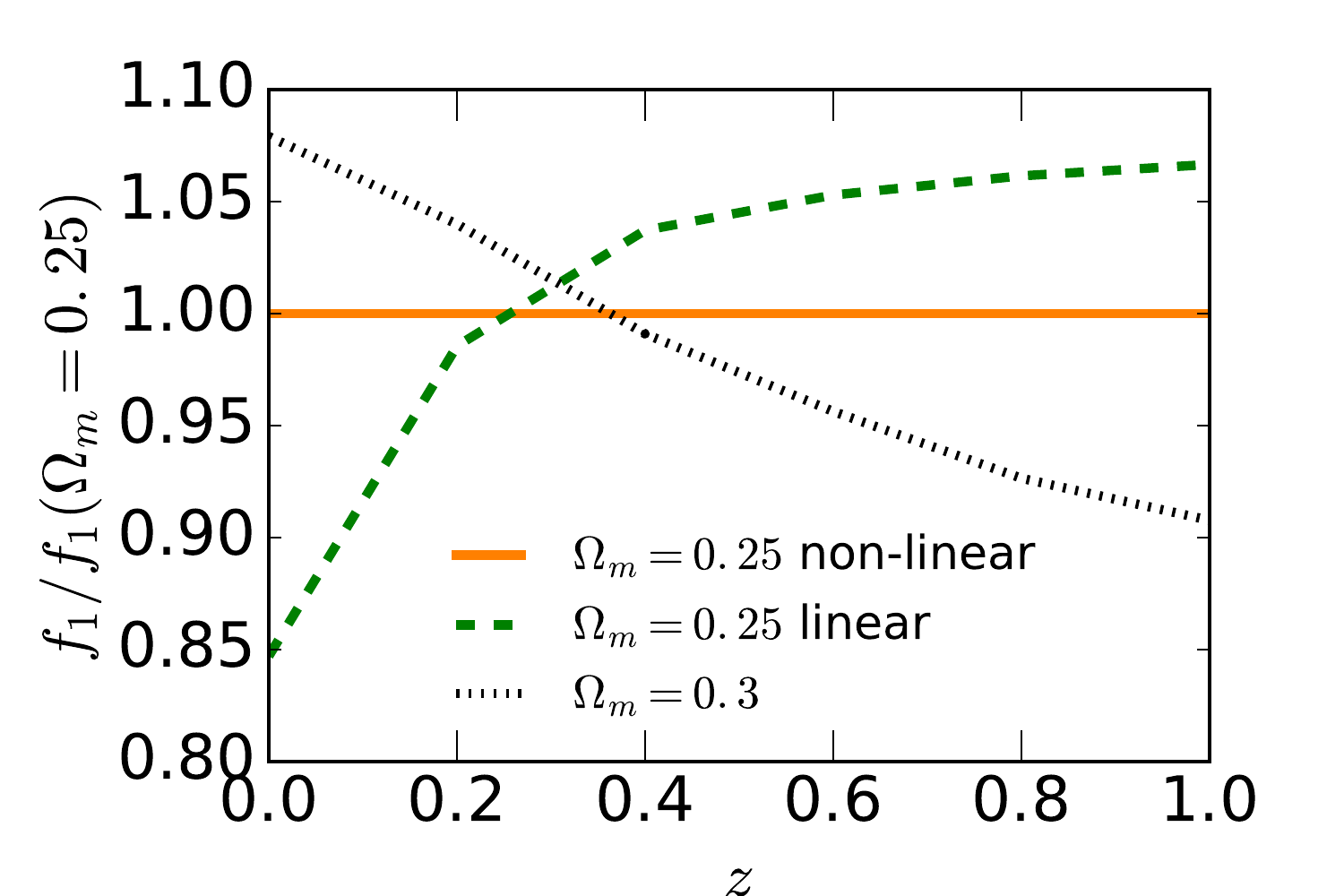}
\includegraphics[scale=.58]{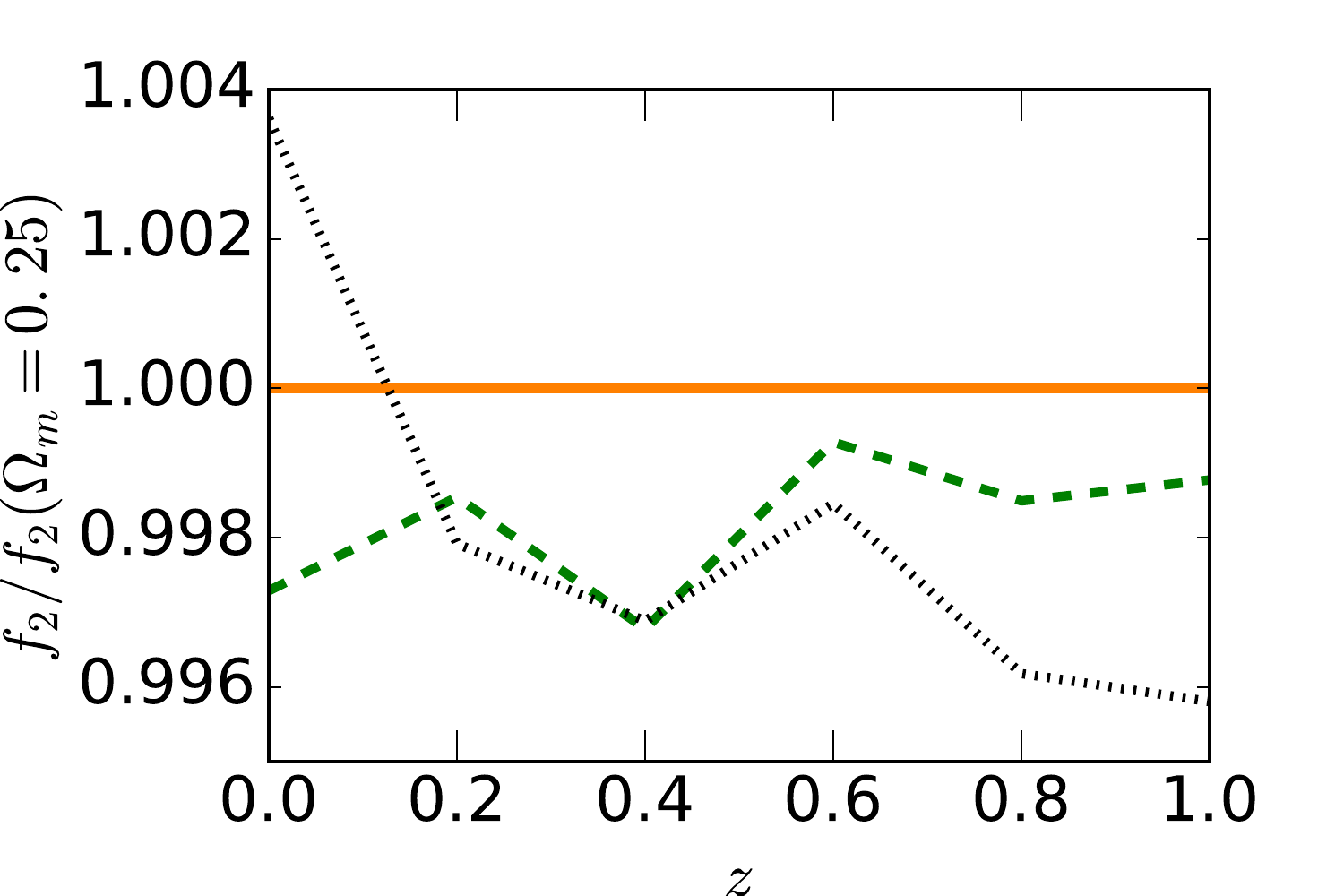}
\caption{$f_1$ (top) and $f_2$ (bottom) for different theory cosmologies, normalized by the values from the MICE cosmology. Dotted lines are obtained for a cosmology with $\Omega_m = 0.3$, while the other two lines represent $\Omega_m = 0.25$. The orange solid line has been obtained using non-linear theory with \textsc{Halofit} \citep{Smith2003}, while the dashed green line has been obtained from linear theory using \textsc{Halofit}.}
\label{fig:f_cosmo}
\end{figure}

Equations (\ref{eq:f_1}-\ref{eq:f_2}) can be predicted theoretically by assuming a cosmology. However, most of the cosmology dependence of the expression is canceled out due to the ratios from $f_1$ and $f_2$, so the final factor is weakly dependent on cosmology. In our case we assume the cosmology of the MICE simulation. In Figure \ref{fig:f_cosmo} we show $f_1$ (top) and $f_2$ (bottom) for different cosmologies and theories, using an angular scale of $50$ arcmin, normalized by the values corresponding to the MICE cosmology. Orange solid lines represent the MICE cosmology, predicted from \cite{Eisenstein1998} non-linear theory obtained using \textsc{Halofit} \citep{Smith2003}. The dashed green lines show the same but obtained from linear theory. We can see that the differences between using linear and non-linear theory are small compared with the final errors that we obtain from our method. We use the old version of \textsc{Halofit} for this prediction, which produces larger differences between the MICE and the theoretical linear Power Spectrum \citep{Fosalba2015}. On the other hand, to obtain the non-linear prediction we computed the non-linearities in an intermediate redshift and extrapolated to the other redshifts using linear growth, which causes a larger disagreement between the linear and non-linear predictions. Because of all this, the difference between the orange and dashed green lines may be interpreted as the upper bound of the disagreement between linear and non-linear theory. 

Finally, in black dotted lines we show the predictions for the same cosmology but with $\Omega_m = 0.3$. We can see that the differences between both cosmologies are smaller than the errors of our bias estimation, even with the fact that the differences in $\Omega_m$ are very large and that $\Omega_m$ is the most sensitive parameter of these predictions. Hence, we can say that the cosmology dependence of this method is very weak.

Equations (\ref{eq:f_1}-\ref{eq:f_2}) describe the contribution of these zero-lag correlations of $\kappa$ and $\kappa'$ in a given redshift bin for the dark matter field. As the dark matter field has a bias of $1$ by definition, using the galaxies instead of the dark matter field to compute $\kappa'_g$ instead of $\kappa'$ in equations (\ref{eq:f_1}-\ref{eq:f_2})  would give $b_{1,2}' f_{1,2}$ instead of $f_{1,2}$ , where $b_{1,2}'$ is the galaxy bias in the redshift bin used (assuming that galaxy bias is constant inside the redshift bin). Then, to estimate galaxy bias in these bins, we need to obtain the bias from equations (\ref{eq:predicted_bias1}-\ref{eq:predicted_bias2}).

\subsection{Numerical effects and parameters}

There are different parameters that can affect our implementation presented in \S\ref{sec:bias_method}. We have studied in which regime our method is valid, or consistent with the linear bias from equation (\ref{eq:bias_2pcf}), and what are the dependences when it is not valid. With this, we can either calibrate our results or restrict to the regimes where our bias measurement is carried out. Here we describe the main numerical effects and our choice of parameters for our final implementation in \S\ref{sec:tomography} and Figure \ref{fig:tomography}. 

\subsubsection*{Catalogue selection}

We used an area of $0^o<RA, DEC <30^o$. This is the same area we used for the fiducial bias measurements from equation (\ref{eq:bias_2pcf}), so that our comparison of both bias is not affected by differences in area or sample variance. This area is similar to DES Y1 data, so this study can be seen as an estimation of the theoretical limitations of this method on DES Y1.

We apply a magnitude cut for the foreground galaxies of $i < 22.5$, to be able to compare it in Paper II with measurements in the DES SV data \citep{Crocce2015b}. However, other selections can be done for this method, such us selecting galaxies by colour or luminosity, in order to measure colour and luminosity dependent bias, that would give information about galaxy formation and evolution. 

\subsubsection*{Redshift bin width}

For the choice of $\Delta z$ we need to take into account two effects. On one side, the use of wide redshift bins would mean losing information from the small scale fluctuations of $\delta_g$ in the line-of-sight, since we project the galaxies in the same bin to measure $\delta_g$. We have seen that this produces a deviation in the value of galaxy bias that is larger than $5\%$ for $\Delta z > 0.2$, and it can be larger than $10\%$ for $\Delta z > 0.3$. We explore this in Figure \ref{fig:bias_vs_Dz} and in \S\ref{sec:testing}. We take this effect into account when we estimate bias in tomographic bins at the end of the paper. When we have photo-z errors, the redshift binning effect is not as important as for the ideal case. If the photo-z errors dominate, the dilution of the small scale fluctuations come from the photo-z errors, and the redshift binning does not affect much. We address the effects of photo-z errors in Paper II. 

On the other hand, the use of narrow redshift bins requires a smoothing of the estimation of $\bar{\rho}_g(z)$. If we calculate $\bar{\rho}_g$ for each redshift bin alone, for narrow bins $\bar{\rho}_g(z)$ is affected by the structure fluctuation in each particular redshift bin, and this causes a smoothing in the final estimation of $\delta_g$. This happens because, when a redshift bin is dominated by an overdensity fluctuation, $\bar{\rho}_g(z)$ is overestimated and hence $\delta_g$ is underestimated. On the other hand, when the redshift bin is dominated by an underdensity, $\bar{\rho}_g(z)$ is underestimated and hence $\delta_g$ is overestimated. The final $\delta_g(z)$ is then smoothed, since all the values tend to be closer to zero due to the calculation of $\bar{\rho}_g(z)$. Some smoothing of $\bar{\rho}_g$ in redshift is needed to avoid this effect when using narrow bins. This is relevant for $\Delta z < 0.03$.

We use redshift bins of $\Delta z = 0.2$ for the foreground galaxies. In this analysis we use the true redshift from the simulation, but in data this method would be also affected by photo-z errors. When photo-z errors are present, using narrow redshift bins is not worth, since the uncertainty in redshift from photo-z errors dominate. We choose this redshift bin width for our estimation of bias in order to test how well we can recover galaxy bias using the redshift binning that is used in Paper II. 

\subsubsection*{Angular scale}

To generate the maps we pixelize the sky using a sinusoidal projection (which consists on redefining $RA$ as $(RA - 15)\cos (DEC)$ in order to obtain a symmetric map with pixels of equal area) with an angular resolution of 50 arcmin, so that the area of the pixels is $(50 \mbox{ arcmin})^2$. Then galaxies are projected in different redshift bins according to their true redshift. 

The bias estimated from this method is not necessarily consistent with the bias from equation (\ref{eq:bias_2pcf}) at small scales. These two methods are only expected to agree at large scales, in the  linear bias regime. Moreover, this method requires a projection in the line-of-sight, so that different scales (weighted differently according to the lensing kernel) are mixed for the same angular scale. However, we have seen that bias is constant for angular scales larger than $\Theta \gtrsim 30$ arcmin, meaning that linear scales are dominant in this regime. In Figure \ref{fig:b2p_vs_bcic} we show the agreement of galaxy bias between equations (\ref{eq:bias_2pcf}) and (\ref{eq:local_bias_corrs1}-\ref{eq:local_bias_corrs3}) when we use a pixel scale of $50$ arcmin, as a visual example of this. 

\subsubsection*{Smoothing}

We do not apply any smoothing in the pixelized maps to estimate galaxy bias in this paper. Exceptionally, for the maps in Figure \ref{fig:ideal_case_constant} we use pixels of $7$ arcmin and we apply a Top Hat filter of $50$ arcmin to smooth the maps. We do this only in this figure in order to have a better visibility of the structures of the maps and the shape of the area used. For the rest of the analysis of the paper, we use pixels of $50$ arcmin and no smoothing kernel afterwards. 



\subsubsection*{Edge effects}

We use a limited area and we project the sky to obtain the maps. When we pixelize the map with a definite pixel scale, due to the projection and the shape of the area used, part of the pixels in the edges are partially affected by these edges. We exclude these pixels from the analysis. 

When a smoothing kernel is applied to the pixelized map, the pixels that are close to the edges are also affected by them. We exclude the pixels whose distance to the edges is smaller than the smoothing radius. 

\subsubsection*{Source redshift and redshift range}

We estimate the $\kappa$ field at $z \simeq 1.3$ by calculating the mean $\kappa$ of the source galaxies with $1.2 < z < 1.4$  in each pixel. The redshift range used ensures we have enough density of galaxies to correctly calculate $\kappa$. 

Theoretically one should take into account the redshift distribution of the source galaxies so that each galaxy contributes to $\kappa_g$ with its position $\chi_s$. However, approximating these galaxies to a plane in their mean position at $z \simeq 1.3$ causes less than a $1\%$ effect.

We use single redshift bins of $\Delta z = 0.2$ for the foreground galaxies in the range of $0.2 < z < 1.2$ to estimate the bias in each of these bins. This produces a galaxy bias estimation of $5$ points in the whole redshift range available (for this method) in the simulation.

 \section{Results}\label{sec:results}

\subsection{Testing}
\label{sec:testing}

In this study we test our method against a fiducial galaxy bias. For this, we measure the angular 2PCFs of matter and galaxies $\omega(\theta)$ and $\omega_g(\theta)$ in the simulation for different redshift bins, using the same area and galaxies that we use for our method. 
We also estimate bias from the definitions in equations (\ref{eq:local_bias_corrs1}-\ref{eq:local_bias_corrs3}) in the same simulation to study the consistency between the different bias definitions.

\begin{figure}
\centering
\includegraphics[scale=.43]{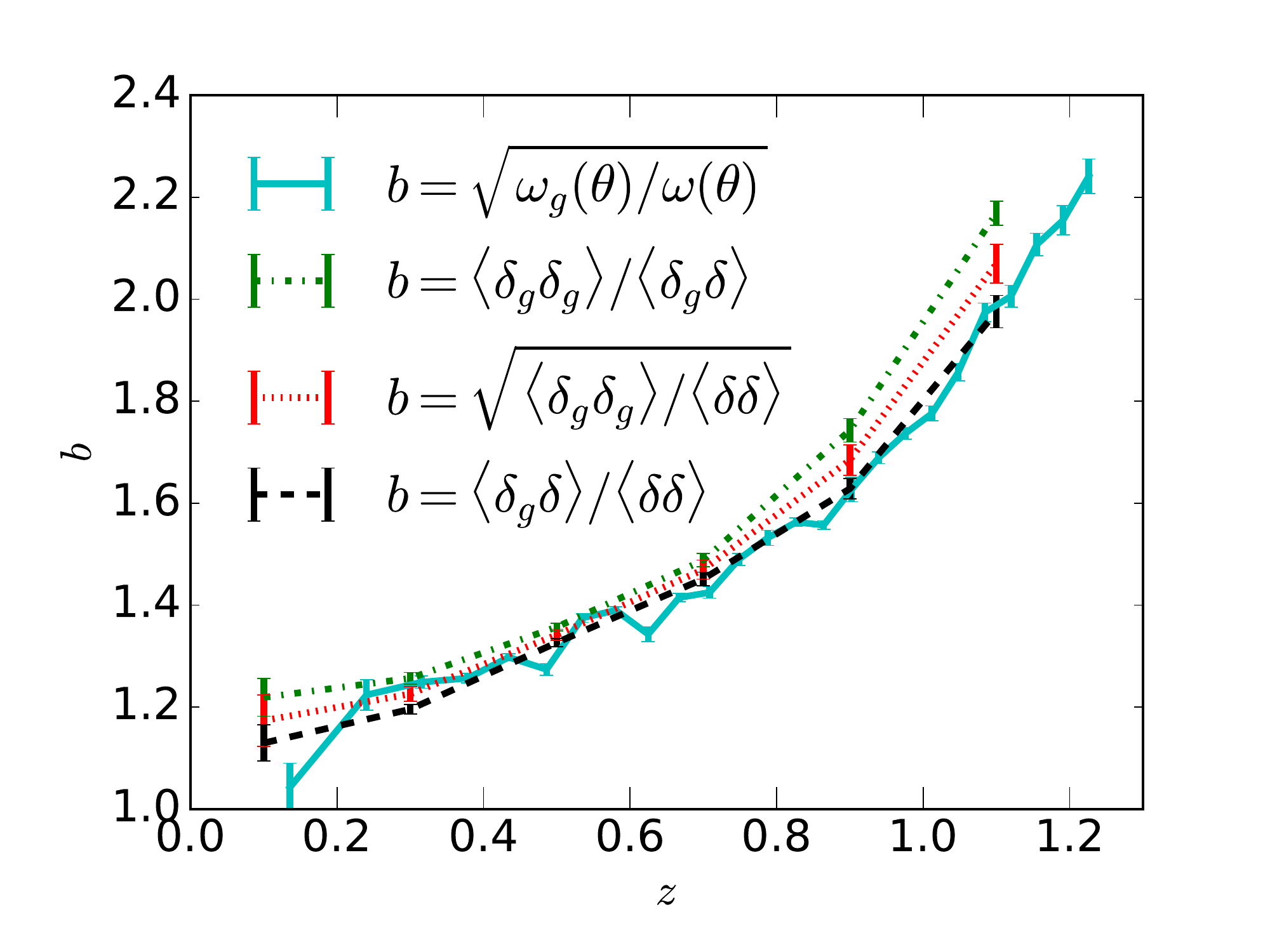}
\caption{Comparison of different definitions of bias. Solid cyan line shows the bias as defined in equation (\ref{eq:bias_2pcf}). The dashed black, dash-dotted green and dotted red lines show bias according to the different definitions from equations (\ref{eq:local_bias_corrs1}-\ref{eq:local_bias_corrs3}). }
\label{fig:b2p_vs_bcic}
\end{figure}

In Figure \ref{fig:b2p_vs_bcic}  we compare different estimators of galaxy bias from the MICE Simulation, using an area of $0^o~<~RA,~DEC~<~30^o$. The solid cyan line represents the bias definition from equation (\ref{eq:bias_2pcf}). We measure $\omega(\theta)$ and $\omega_g(\theta)$ as a function of the angular scale, and to obtain the bias we fit the ratio as constant between $6$ and $60$ arcmin. The angular correlation function involves different comoving scales for different redshifts, and then fixing the same angular scales for the galaxy bias implies a mix of physical scales. However, for large enough scales, bias is constant and is not affected by this. We have checked that bias does not change significantly at these scales, and using these scales to measure galaxy bias give consistent results with using larger scales. The galaxy bias obtained from equations (\ref{eq:local_bias_corrs1}-\ref{eq:local_bias_corrs3}) are shown in dashed black line, dotted red line and dash-dotted green line as specified in the legend. This has been calculated in each redshift bin by pixelating $\delta$ and $\delta_g$ in pixels of area $(50 \mbox{ arcmin})^2$ using redshift bins of $\Delta z = 0.2$. The agreement between the solid cyan and the dashed black lines confirms that linear bias from equation (\ref{eq:bias_2pcf}) is consistent with local bias measure from equation (\ref{eq:local_bias_corrs1}) at these scales. On the other hand, the differences in the different expressions of  equations (\ref{eq:local_bias_corrs1}-\ref{eq:local_bias_corrs3}) implies a stochasticity between $\delta_g$ and $\delta$ that affects our estimations of bias. We see that the same effect appears when using equation (\ref{eq:zero_lag_correct2}) to estimate galaxy bias, and this can be explained by the projection effect due to the redshift binning, as discussed below in Figures \ref{fig:bias_vs_par} and \ref{fig:bias_vs_Dz}. We take into account this effect to estimate tomographic bias in \S\ref{sec:tomography}.

The idea of the following analysis is to test how the calculations of this method deviate from the expected estimation of linear bias using different angular scales and binning. 
For these testing purposes, we construct here the bias-corrected $\kappa_{g}$ map, $\hat{\kappa}_{g}$, defined as:
\begin{equation}
\hat{\kappa}_g(b, \bm\theta) = \sum_{i = 1}^N{q^i \frac{\delta^i_g(\bm\theta)}{b^i}\Delta\chi^i},
\label{eq:kappa_int_bias}
\end{equation}
where $b^i$ corresponds to the linear bias measured in $N$ bins that can be obtained from equations (\ref{eq:bias_2pcf}) or (\ref{eq:local_bias_corrs1}-\ref{eq:local_bias_corrs3}).
In analogy with equations (\ref{eq:zero_lag1},\ref{eq:zero_lag2}), we can calculate the corresponding normalized bias between the $\hat{\kappa}_{g}$ and $\kappa$ fields:
\begin{equation}
\hat{b}(b) = \frac{\langle \hat{\kappa}_g(b) \kappa \rangle}{\langle \kappa \kappa \rangle - \langle \kappa^N \kappa^N \rangle} 
\label{eq:zero_lag_correct1}
\end{equation}
\begin{equation}
\tilde{b}(b) = \frac{\langle \hat{\kappa}_g(b) \hat{\kappa}_g(b) \rangle - \langle \kappa_g^N \kappa_g^N\rangle}{\langle \hat{\kappa}_g(b) \kappa\rangle}.
\label{eq:zero_lag_correct2}
\end{equation}
Note that $\hat{b}$ and $\tilde{b}$ depend on the bias $b$ used to obtain $\hat{\kappa}_g$. Under this definition, $\hat{b} = 1$ and $\tilde{b} =1$ suggest that this method is consistent with measuring linear bias $b$.

\begin{figure}
\centering
\includegraphics[scale=.43]{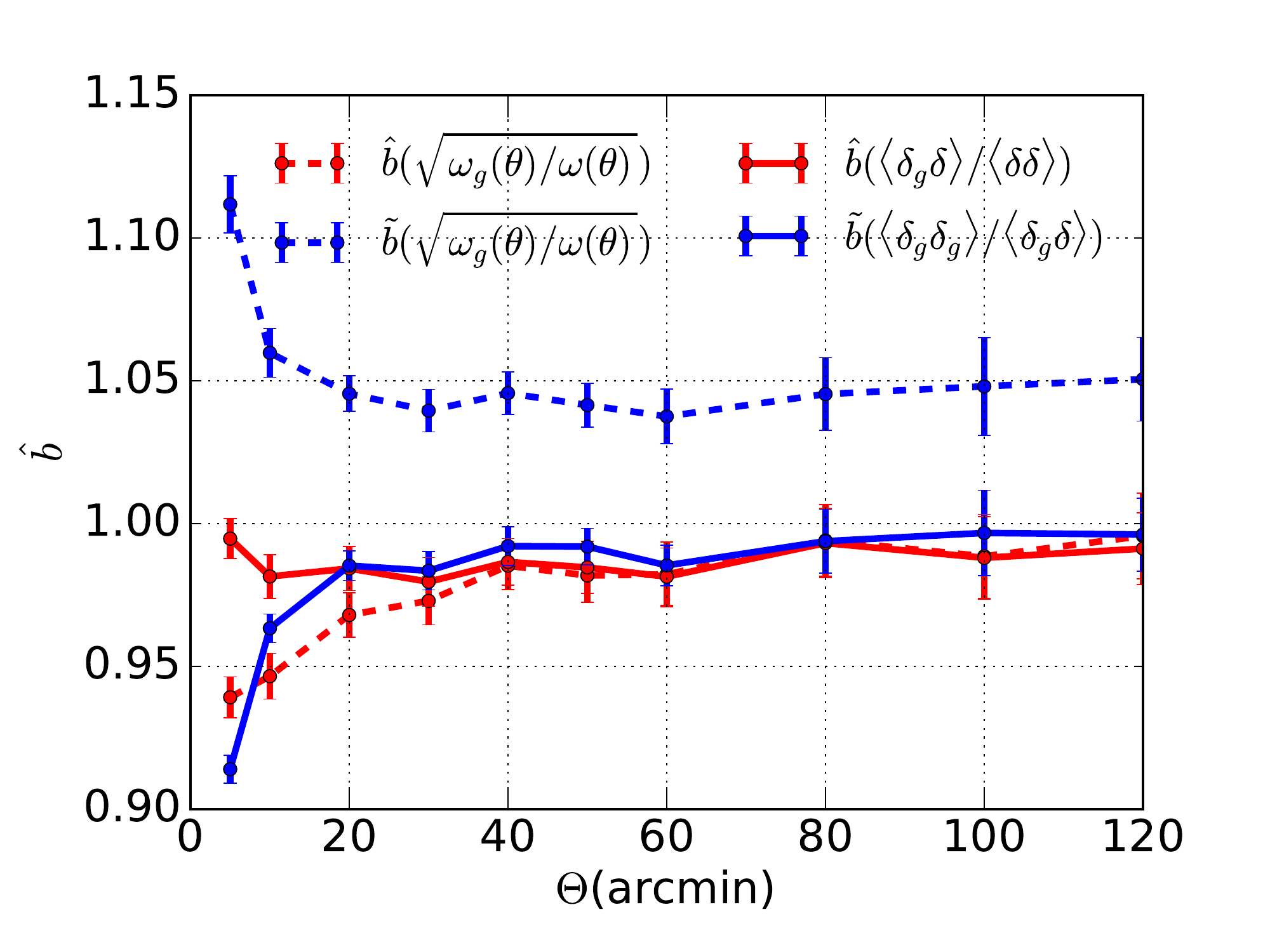}
\caption{Bias from the zero-lag cross correlations of $\kappa$ and $\hat{\kappa}_{g}$ as a function of angular scale, where $\hat{\kappa}_g$ is an estimation of $\kappa_g$ normalized by the redshift dependent bias from different estimators as in equations (\ref{eq:kappa_int_bias}-\ref{eq:zero_lag_correct2}). Dashed red line shows $\hat{b}(\sqrt{\omega_g(\theta)/\omega(\theta)})$. The solid red line shows the same, but obtaining $\hat{\kappa}_g$ from the bias from equation (\ref{eq:local_bias_corrs1}) to obtain $\hat{b}(\langle \delta_g \delta \rangle / \langle \delta \delta \rangle)$. The dashed blue line shows $\tilde{b}(\sqrt{\omega_g(\theta)/\omega(\theta)})$. The solid blue line shows the same $\tilde{b}$, but normalizing $\hat{\kappa}_g$ from equation (\ref{eq:local_bias_corrs2}) to obtain $\tilde{b}(\langle \delta_g \delta_g \rangle / \langle \delta_g \delta \rangle)$.}
\label{fig:bias_vs_par}
\end{figure}

Figure \ref{fig:bias_vs_par} shows how the estimators $\hat{b}$ and $\tilde{b}$ change as a function of the angular scale, defined by the pixel scale, for different estimators of bias used to obtain $\hat{\kappa}_g$.  For the dashed red and blue lines we used $b(z)$ from equation (\ref{eq:bias_2pcf}) to obtain $\hat{\kappa}_g$ for our estimation of $\hat{b}(\sqrt{\omega_g(\theta)/\omega(\theta)})$ and $\tilde{b}(\sqrt{\omega_g(\theta)/\omega(\theta)})$ respectively.
We can see that the measurements are constant for $\Theta > 30$ arcmin, meaning that we are in the linear regime in these scales. However, there is a $5\%$ difference between the two estimators at large scales (at small scales nonlinearities appear and the difference is larger). This can be interpreted from Figure (\ref{fig:b2p_vs_bcic}), where we see that the estimators from equations (\ref{eq:local_bias_corrs1}) (represented as a dashed black line) and (\ref{eq:local_bias_corrs2}) (represented as a dash-dotted green line) are slightly different. In fact, $\hat{b}$ is indirectly measuring equation (\ref{eq:local_bias_corrs1}), which is consistent with bias from equation (\ref{eq:bias_2pcf}) (at the $1\%$ level), while $\tilde{b}$ is indirectly measuring equation (\ref{eq:local_bias_corrs2}), which is slightly higher than bias from equation (\ref{eq:bias_2pcf}). If we use equation (\ref{eq:local_bias_corrs1}) for the calculation of $\hat{\kappa}_g$ to obtain $\hat{b}(\langle \delta_g \delta \rangle / \langle \delta \delta \rangle)$ (shown in the solid red line) and equation (\ref{eq:local_bias_corrs2}) for the calculation of $\hat{\kappa}_g$ to obtain $\tilde{b}(\langle \delta_g \delta_g \rangle / \langle \delta_g \delta \rangle)$ (shown in the solid blue line), then both estimations are consistent, as expected. As in Figure \ref{fig:b2p_vs_bcic}, the difference between both estimators coming from $\hat{b}(\sqrt{\omega_g(\theta)/\omega(\theta)})$ and $\tilde{b}(\sqrt{\omega_g(\theta)/\omega(\theta)})$ can be seen as an indication (and a measurement) of the stochasticity in the relation between $\delta_g$ and $\delta$, giving a factor of $5\%$.

In order to go deeper in the analysis of these effects and see whether these differences between both estimators come from the intrinsic relation between $\delta_g$ and $\delta$ or from numerical systematics, we constructed the following template $\kappa_m$:
\begin{equation}
\kappa_m (\bm\theta) = \sum_{i = 1}^N q^i \delta^i (\bm\theta) \Delta \chi^i,
\end{equation}
which corresponds to the same exact calculation than equation (\ref{eq:kg_sum}) for $\kappa_g$, but using dark matter particles instead of  galaxies. This field $\kappa_m$ is expected to reproduce $\kappa$ from the Born approximation consistently except for the numerical differences between the method and how the original $\kappa$ is obtained, which basically come from the redshift binning and projection discussed below equation (\ref{eq:kg_sum}). In order to avoid noise in the $\kappa$ map, we use $\kappa_T$, defined as the true map directly obtained from the high resolution map of the simulation \citep[see ][]{Gaztanaga1998,Fosalba2008,Fosalba2015b}, and we calculate the bias of these two estimators of $\kappa$ as:
\begin{equation}
b_{m,1} = \frac{\langle \kappa_m \kappa_T \rangle}{\langle \kappa_T \kappa_T \rangle} 
\label{eq:zero_lag_dm1}
\end{equation}
\begin{equation}
b_{m,2} = \frac{\langle \kappa_m \kappa_m \rangle - \langle \kappa_m^N \kappa_m^N \rangle}{\langle \kappa_m \kappa_T \rangle},
\label{eq:zero_lag_dm2}
\end{equation}
that should give $b_{m,1,2} = 1$ if there are no numerical systematics. 

\begin{figure}
\centering
\includegraphics[scale=.43]{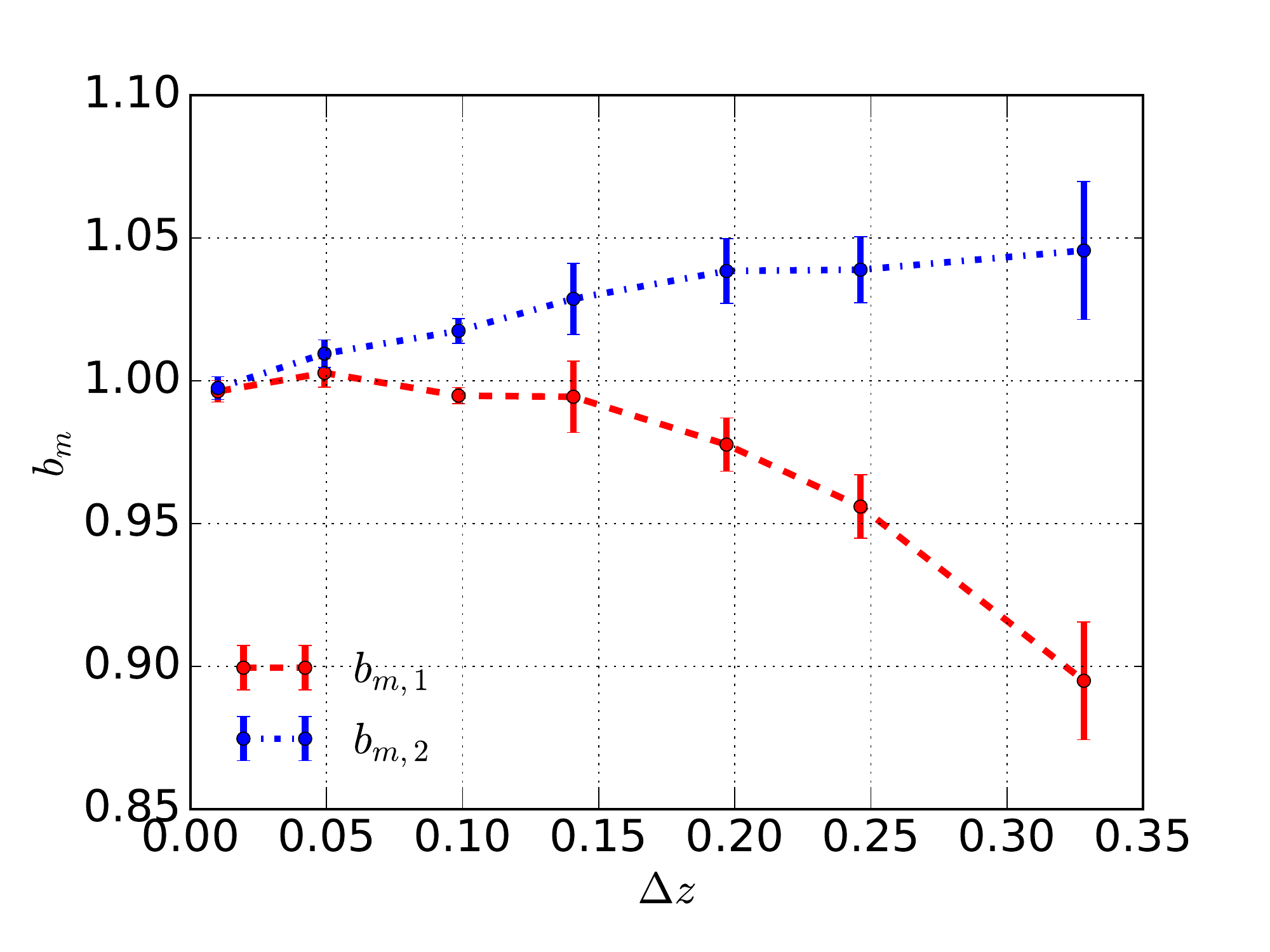}
\caption{$b_m,1$ and $b_{m,2}$, defined in equations (\ref{eq:zero_lag_dm1},\ref{eq:zero_lag_dm2}), as a function of the redshift bin width used, $\Delta z$, for the two estimators.}
\label{fig:bias_vs_Dz}
\end{figure}

We have found that $b_{m,1,2}$ behaves as $\hat{b}$ and $\tilde{b}$ in our tests, meaning that the differences between the different estimators can be seen as a measurement of the numerical effects on the method. In fact, we have found that the differences mainly come from the projection effect in the redshift bins, as shown in Figure~\ref{fig:bias_vs_Dz}. Here we show $b_{m,1}$ and $b_{m,2}$ as a function of the redshift bin width, $\Delta z$, used to obtain $\kappa_m$. We use a pixel scale of $50$ arcmin, a source redshift of $z_s = 1$ and we use all the dark matter particles (diluted with respect to the total number of particles, but this does not affect the result) within $z < 1$. We see that the two estimators agree when we use narrow redshift bins, but the difference between both increases with $\Delta z$. For $\Delta z = 0.2$, the difference is the $5\%$ that we see in Figure \ref{fig:bias_vs_par} for the galaxies. This test measures the redshift binning and the projection impacts on this method, and can also be used to calibrate the measurements. In fact, $f_1$ and $f_2$ can be used to take into account these projections, specified by the selection function $p'(\chi)$, and the redshift binning.
But  in the case of Figures \ref{fig:bias_vs_par} and \ref{fig:bias_vs_Dz}, we use all the redshift range in the foreground of the sources and we have not corrected by $f_1$ and $f_2$.
In the case of Figure \ref{fig:bias_vs_par}, this effect is visible for $\hat{b}(\sqrt{\omega_g(\theta)/\omega(\theta)})$ and $\tilde{b}(\sqrt{\omega_g(\theta)/\omega(\theta)})$. However, for $\hat{b}(\langle \delta_g \delta \rangle / \langle \delta \delta \rangle)$ and $\tilde{b}(\langle \delta_g \delta_g \rangle / \langle \delta_g \delta \rangle)$ the projection effect is compensated because we use the same redshift binning (and then the projection effects are the same and compensate) to obtain $\langle \delta_g \delta \rangle / \langle \delta \delta \rangle$, $\langle \delta_g \delta_g \rangle / \langle \delta_g \delta \rangle$, $\hat{b}$ and $\tilde{b}$.
In the next section we will apply the $f_1$ and $f_2$ corrections to the tomographic estimations.

\subsection{Redshift dependence of bias}\label{sec:tomography}

\begin{figure}
\centering
\includegraphics[scale=.43]{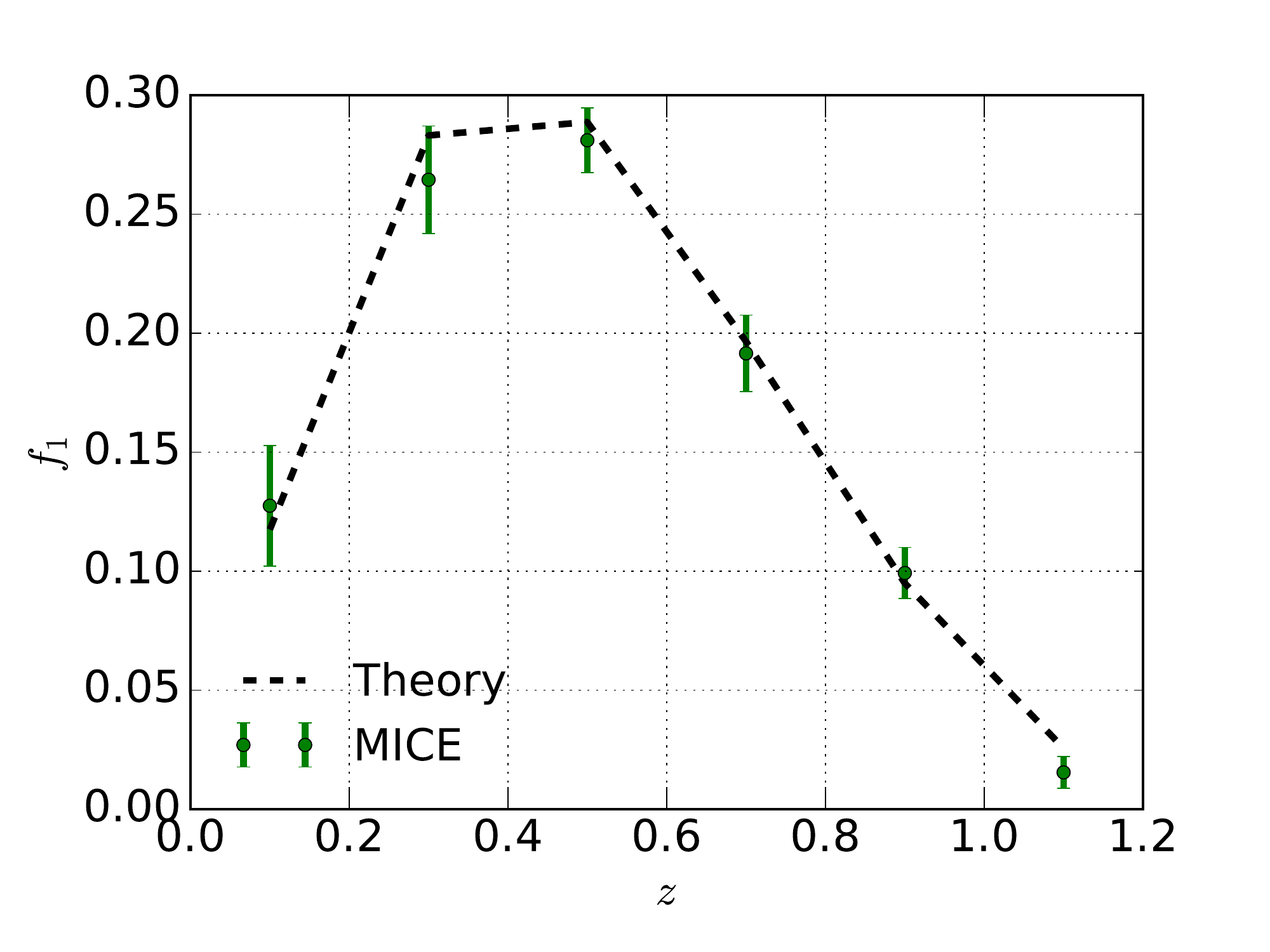}
\caption{Comparison of $f_1$ from the simulation (green points) and the theory prediction (dashed black line). Each value has been obtained by using redshift bins of $\Delta z = 0.2$ to calculate $\kappa'$, and using a source redshift of $z_s = 1.3$.}
\label{fig:z_range}
\end{figure}

In Figure \ref{fig:z_range} we show a comparison between the theoretical predictions (in dashed black lines) of $f_1$ and the measurements in the MICE simulation (in green points), in 6 different redshift bins of $\Delta z = 0.2$, using a redshift for the sources of $z_s = 1.3$.  We see a good agreement between theory and simulations. The redshift dependence of $f_1$ comes from the contributions of the lensing kernel, that causes the amplitude of $f_1$ to be higher at intermediate redshifts, but is also affected by the binning (that implies a projection of $p'(\chi)\delta(\chi)$ inside the bin) and the correlation functions from equations (\ref{eq:kprime_k}-\ref{eq:k_k}) (that has a contribution coming from the correlation between the dark matter distribution inside and outside the bin).
The redshift dependence of the amplitude of $f_1$ reflects the contribution to $\hat{b}$ of each of these redshift bins.

Equations (\ref{eq:predicted_bias1},\ref{eq:predicted_bias2}) give a tool that can be used for tomographic measurements of galaxy bias, since we can estimate the bias using different redshift bins of the foreground galaxies if we take this correction into account. That is, we can measure $b_{1,2}'$ for a given redshift by calculating $\kappa'_g$ in that bin and using equations (\ref{eq:predicted_bias1},\ref{eq:predicted_bias2}). 
 
\begin{figure}
\includegraphics[scale=.43]{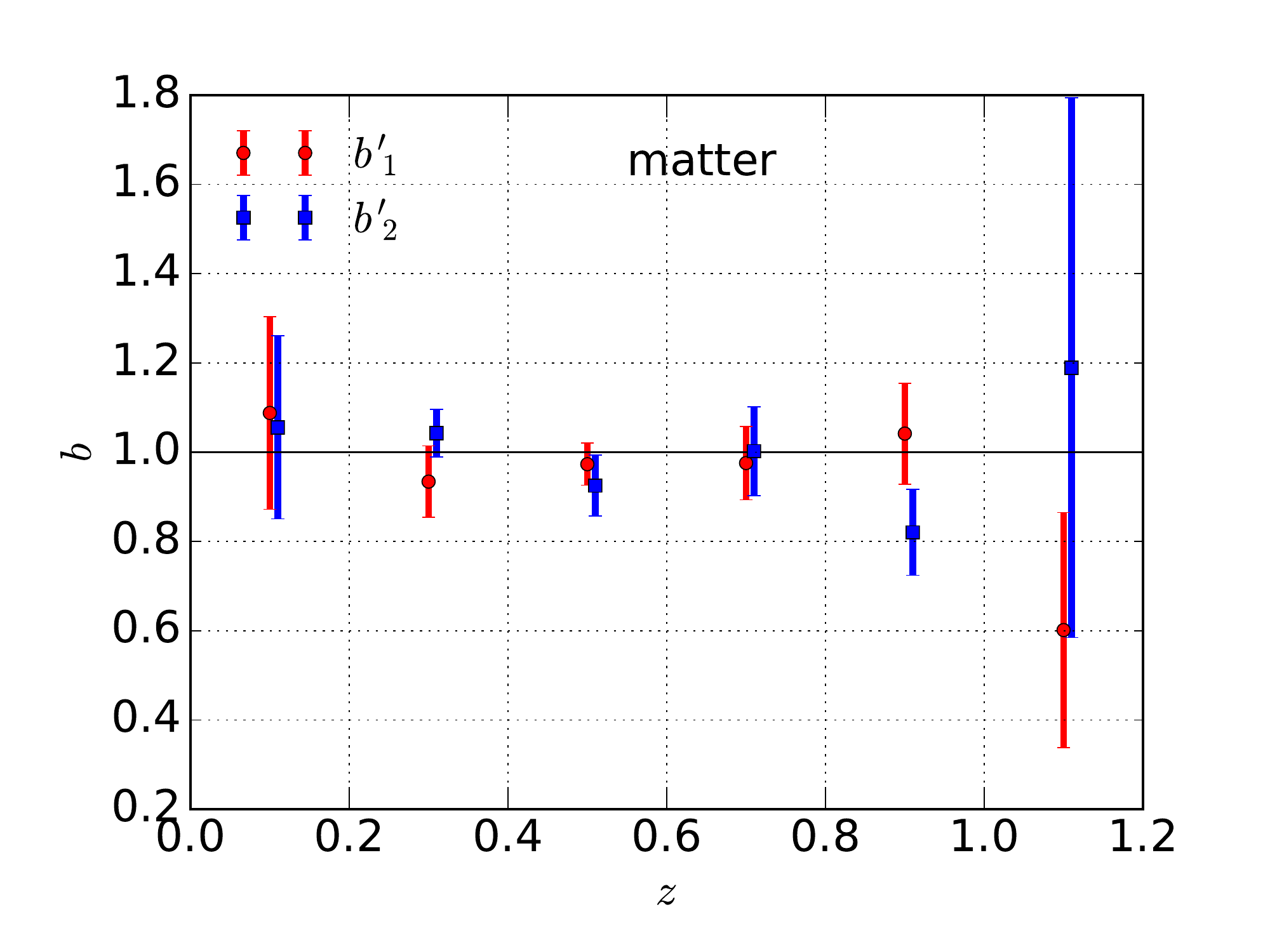}
\caption{Tomographic bias of dark matter in the MICE simulation, using the two bias estimators from equations (\ref{eq:predicted_bias1},\ref{eq:predicted_bias2}). We use tomographic redshift bins of $\Delta z = 0.2$ and a source redhift of $z_s = 1.3$. }\label{fig:tomography_matter}
\end{figure}

In Figure \ref{fig:tomography_matter} we show a test where we obtain tomographic bias of dark matter from the MICE simulations using this method. For this, we use $\kappa'$ instead of $\kappa'_g$ in equations (\ref{eq:predicted_bias1},\ref{eq:predicted_bias2}) in order to estimate the bias of matter. By construction, the results should be consistent with $1$. Our results are consistent, meaning that our method estimates tomographic bias correctly. We observe that the errors of these estimations are smaller in the intermediate redshifts and larger at the extremes. This is due to the redshift dependence of $f_1$, that optimizes the signal for a high amplitude of $f_1$. When $f_1$ is small, the estimation of bias becomes noisier.

\begin{figure}
\includegraphics[scale=.43]{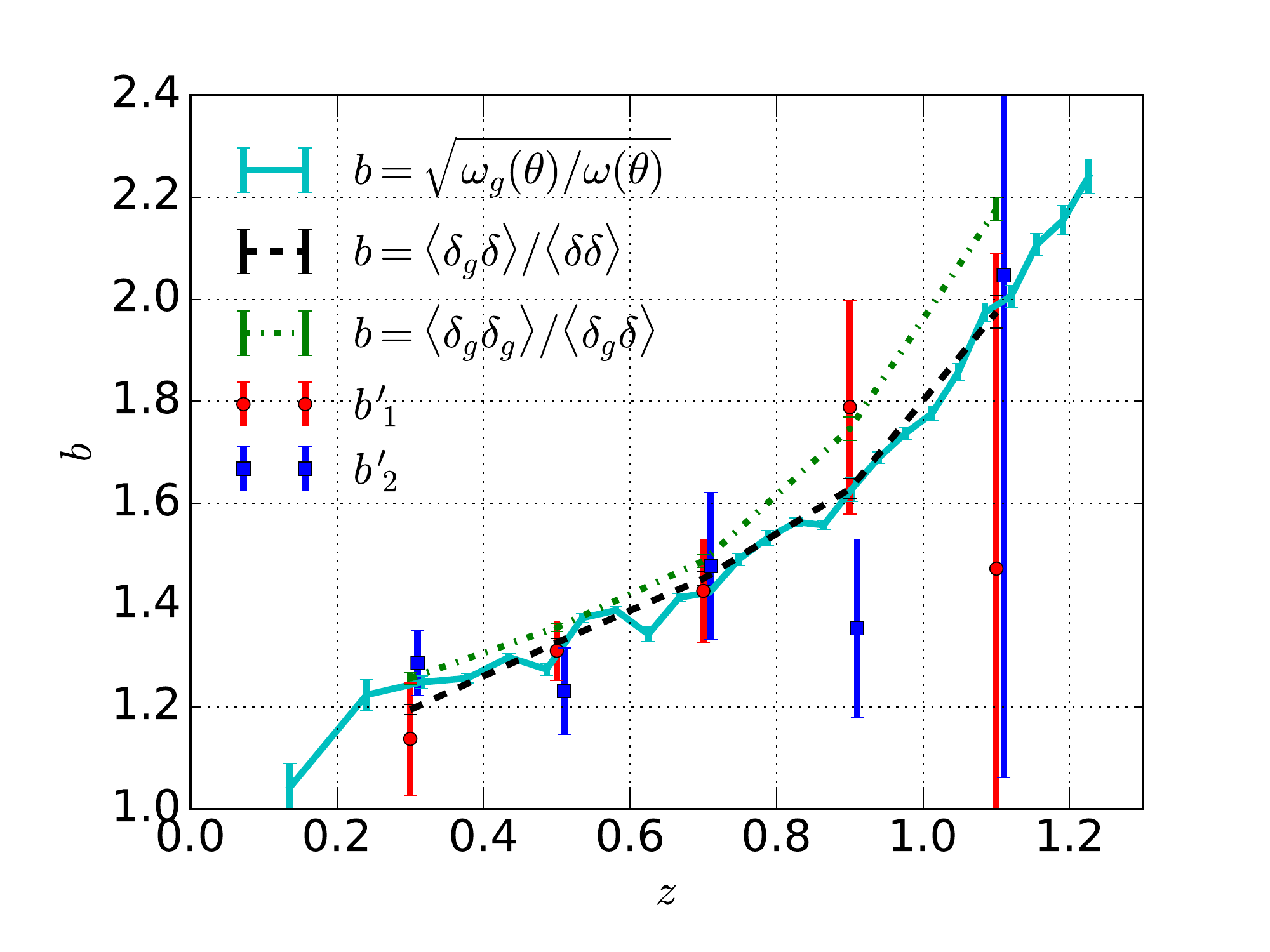}
\caption{Redshift dependent bias estimated from our method, shown in red and blue points as specified in the legend and equations (\ref{eq:predicted_bias1}-\ref{eq:predicted_bias2}). For this we used tomographic redshift bins of $\Delta z = 0.2$ and a source redshift of $z_s = 1.3$. The solid cyan line shows linear bias from equation (\ref{eq:bias_2pcf}), fitting bias as constant between $6$ and $60$ arcmin. The dashed black line shows bias estimated from $\langle \delta_g \delta \rangle / \langle \delta \delta \rangle$, using the same redshift bins of $\Delta z = 0.2$. The dash-dotted green line shows bias estimaed from $\langle \delta_g \delta_g \rangle / \langle \delta_g \delta \rangle$ in the same redshift bins.}\label{fig:tomography}
\end{figure} 
 
Figure \ref{fig:tomography} shows the estimation of the tomographic galaxy bias using different redshift bins of $\Delta z = 0.2$ for both estimators from equations (\ref{eq:predicted_bias1},\ref{eq:predicted_bias2}), represented as red and blue points as specified in the legend. We compare them with the fiducial bias from equations (\ref{eq:bias_2pcf},\ref{eq:local_bias_corrs1},\ref{eq:local_bias_corrs2}) shown in solid cyan, dashed black and dash-dotted green lines respectively. We see that the method we present in this paper gives consistent results with linear bias. There are some slight differences for the estimator from equation (\ref{eq:local_bias_corrs2}) which, as mentioned above, is due to the effects of projection and binning. But this effect is not shown from the tomographic bias obtained from our method, because we take into account this effect in the factors $f_1$ and $f_2$. Note also that the two methods, represented by the red and blue points, give very similar results (apart from the fourth bin). 

We can see that the errors are very large for the highest redshift bin. This is due to the fact that, due to the lensing kernel, $f_1$ and $f_2$ are very small, and then the measurements in this bin are very sensitive to small changes. The best error bars appear where the lensing kernel is higher, so the potential of this method is optimal in the maximum of the lensing kernel. Hence, different source redshifts might be combined in order to optimize the analysis for all redshifts. In Paper II we combine the results using multiple redshift bins for the source galaxies, and we fit the galaxy bias from the combination of these measurements, using both $\kappa_g$ and $\gamma_g$ and doing a full-covariance analysis. In this paper we do not apply any fit, since we directly measure bias from equations (\ref{eq:predicted_bias1},\ref{eq:predicted_bias2}) using a fixed source redshift bin.



In this paper we show this method for the most idealistic case. However, when applying this method to observations other effects appear to be relevant. First of all, we cannot measure $\kappa$ directly from observations, and then we need to convert $\gamma$ to $\kappa$ or $\kappa_g$ to $\gamma_g$ to measure bias from equations (\ref{eq:zero_lag1},\ref{eq:zero_lag2}) or (\ref{eq:zero_lag_gamma1},\ref{eq:zero_lag_gamma2}). This involves some edge effects when these conversions are applied to a finite area. Moreover, foreground galaxy incompleteness, photo-z estimation, shape noise, mask and intrinsic alignments can affect our results in observations. We address these effects in Paper II.

\section{Discussion and Conclusions}\label{sec:conclusions}

In this paper we explore a new method to measure galaxy bias from the combination of the galaxy density and weak lensing fields. This method is based on A12, where they use the galaxy density field to construct a bias-weighted convergence map $\kappa_g$ in the COSMOS field. They measure different parameterizations of galaxy bias from the zero-lag correlations of the galaxy shear and a reconstruction of the shear from the galaxy density field. In this paper  we present a new way to measure tomographic bias from the zero-lag correlations between the lensing maps and a reconstruction of the lensing maps from the galaxy density field. We also study the robustness and the systematics of this method for the first time. 

The implementation of this model is as follows. We construct a template of the convergence field $\kappa_g$ at the source redshift by integrating the density field of the foreground galaxies in the line-of-sight weighted by the corresponding lensing kernel as specified in equation (\ref{eq:kg}). We do this for tomographic bins in the lens distribution to obtain $\kappa'_g$ as defined in equation (\ref{eq:partial_kg}).
We then compare to estimates of the matter convergence map $\kappa$  associated to the same galaxies in the source redshift bin. 
We measure galaxy bias from the  zero-lag cross-correlations between $\kappa$ and $\kappa'_g$ as in equations (\ref{eq:predicted_bias1},\ref{eq:predicted_bias2}).
Instead of using the zero-lag cross-correlation we could also use the 2-point cross-correlation function. We will apply this for DES Y1 data in a follow-up paper. 

We use the MICE simulations to study the consistency of our method by comparing our results with a fiducial galaxy bias measurement on linear scales. This is obtained from the ratio between the projected 2PCFs of galaxies and dark matter as a function of redshift (see equation (\ref{eq:bias_2pcf})), and fitting a constant galaxy bias between $6$ arcmin and $60$ arcmin. We also study local bias from equations (\ref{eq:local_bias_corrs1}-\ref{eq:local_bias_corrs3}), making use of the dark matter field of the simulation. With these comparisons we study the systematics of the method and the regimes where it is consistent with linear bias.

There are different systematic effects and numerical dependencies of the method that need to be taken into account for a correct measurement of linear bias. First of all, the method is sensitive to the redshift bin width used in the construction of $\kappa_g$ and $\kappa'_g$, that have an impact on the galaxy bias estimators due to the projection effects of the density fields. This causes differences in the values obtained for the different estimators, that can be larger than $5\%$ for $\Delta z > 0.2$ and larger than $10\%$ for $\Delta z > 0.3$.  This has to be taken into account and corrected when measuring $\kappa_g$ and $\kappa'_g$ in wide redshift bins in order to obtain the correct linear bias. On the other hand, assuming that all the source galaxies (selected in a redshift bin of $\Delta z > 0.2$) are in a plane have an insignificant impact on the results. Secondly, the angular scale of the field can be affected by nonlinearities for small enough scales. We find that the measurements are consistent with linear bias for angular scales of $\Theta > 30$ arcmin, where bias is constant. Sampling and discreteness noise is also important and needs to be taken into account (see equations (\ref{eq:zero_lag1},\ref{eq:zero_lag2})).
Finally, we need to exclude from the analysis those pixels that are affected by the edges of the area used.

The true $\kappa$ field comes from the contribution of all the mass distribution in the whole redshift range below the source redshift. Then, if we only use a fraction of this redshift range to calculate $\kappa_g$, the correlation between $\kappa$ and $\kappa_g$ is lower due to the fact that we are not comparing the same redshift ranges. Then, a correction must be applied to our estimators if we only use the foreground galaxies in a given redshift bin for the construction of $\kappa_g$.  We predict theoretically this effect, and we find good agreement with the measurements, indicating that we can use this prediction to correct the bias obtained. The theoretical prediction describes the amplitude of the zero-lag correlations obtained using a given redshift bin for the foreground dark matter field, that by definition has a bias of $1$. We can measure galaxy bias in that bin from the ratio between the zero-lag correlations of $\kappa$ and $\kappa_g$ (using the foregroud galaxies in that bin) and the theoretical prediction as described in equations (\ref{eq:predicted_bias1},\ref{eq:predicted_bias2}). This provides a useful tool to do tomography and measure galaxy bias in single redshift bins. We measure and show the redshift-dependent bias obtained using this method for a flux-limited galaxy sample of $i<22.5$, and find good agreement with the redshift-dependent bias from equation (\ref{eq:bias_2pcf}). 

Other issues associated with observational data must be addressed if we apply this method to large galaxy surveys such as the Dark Energy Survey (DES). These issues include the conversion from $\kappa_g$ to $\bm\gamma_g$ or from $\bm\gamma$ to $\kappa$ and its border effects, shape noise, masking, galaxy incompleteness, photo-z errors and boundary effects. As we measure fluctuations in the convergence maps we are not affected by the mass-sheet degeneracy.  We do not expect intrinsic alignments to affect our results for several reasons. First, we use wide redshift binning for the cross correlations between $\kappa_g$ (or $\bm\gamma_g$) and $\kappa$ (or $\bm\gamma$) at different redshifts. Second, we never correlate lensing maps at different redshifts, so there is no \cite{Seljak2004b} effect in our method. Finally, we could have intrinsic alignment effects from the correlations of lensing maps at the same redshift, so from $\langle \kappa \kappa \rangle$ or $\langle \bm\gamma\bm\gamma\rangle$. However, the estimators that optimize shape noise do not make use of these correlations, so the estimators that we want to use in observations, as applied in \cite{Chang2016}, do not present these contributions. 
We apply this method to the DES Science Verification data in a follow-up paper \citep{Chang2016}. This method is expected to be significantly better when applied to larger areas, such as in DES Year 1 \citep{Diehl2014} or the $5000\mbox{ deg}^2$ from the expected total area of the DES survey, since the statistical errors will be smaller. 

This paper presents the method, but further studies can be done. We can explore galaxy bias for different galaxy samples, e.g. as a function of colour and luminosity. We expect different values of galaxy bias for different properties, since the clustering of galaxies depends on their properties \citep[e.g.][]{Zehavi2011}. However, the accuracy of the method can also depend on the galaxy selection. First of all, the number density of galaxies has an impact on the precision of our estimation of the density field. Because of this, using a high threshold in colour or luminosity would imply a larger uncertainty in the measurement. Moreover, environmental dependencies of galaxy bias might cause a stochasticity between galaxies and matter that might affect the bias estimation. This can happen for old and red galaxies, that are sensitive to environment \citep{Paranjape2015,Pujol2015}. We can also explore the scale dependence of local bias by studying different angular scales and its nonlinearities, and the redshift dependence by comparing the tomographic measurements with parametric redshift-dependent bias based on A12. 
In this paper we have focused on zero-lag cross-correlations, but we could also use 2-point cross-correlations as a way to estimate the bias
and include the redshift cross-correlations as a validation test.

The method studied in this paper has several attractive features. First of all, the method is weakly dependent on the cosmological parameters (it only depends strongly on $\Omega_m$, as many weak lensing measurements). It depends very weakly on $\sigma_8$ (only in non-linear corrections
to $f_1$ and $f_2$), while other measurements of bias (from clustering statistics for example) are typically strongly dependent on $\sigma_{8}$ \citep{Crocce2015b,Giannantonio2016}. This is because the ratios in the $f_1$ and $f_2$ factors cancel out most of the cosmology dependence. These factors include the 2PCFs and geometric distances, that assume a flat Universe and geometry. Another advantage of the method is that it makes use of the lensing maps to measure the matter distribution. Galaxy bias comes from the direct comparison between the galaxy and matter distribution, while other methods usually compare the galaxy distribution observed with a simulated matter distribution. The method is also a good complement to other methods to measure galaxy bias from weak lensing, such as from galaxy-galaxy lensing or from cross-correlations between apperture mass and number counts. Apart form the fact that the cosmology dependences are different, and hence a good complement, our method allows to study local bias in the closest way to theory and N-body estimations. In that sense, our method is the most direct way to measure local bias with observations of both dark matter and galaxies. 

This method is also similar to \cite{Hoekstra2002} and other measurements of bias from the cross-correlation between aperture mass and number counts. However, \cite{Hoekstra2002} look at small scales, where bias shows a significant scale dependence. In our case, we focus on large scales, where bias is not scale dependent and it is consistent with linear bias. In fact, we have tested at what scales our measurement is consistent with linear bias and applied the method to these scales. Moreover, \cite{Hoekstra2002} uses the aperture mass statistics corresponding to a smoothing kernel of the matter field which is different than the one we use here, and A12 showed that different smoothing schemes can produce different values of bias (due to the different contributions of the smallest scales). Our method allows the application of any smoothing scheme that might be useful for any particular study, although we use a box car smoothing that is easily comparable to bias from N-body simulations and theory. Hence, from our method we can study both linear and nonlinear bias by using large or small scales, although here we focus on linear scales. 

Hence, a combined analysis of different measurements of galaxy bias, including this method, can be very useful to constrain better bias and cosmology. The method can also be applied to a situation where galaxies only cover
partially the full redshift range of the lenses.
Finally, the potential of this method will rapidly increase with the data of present and upcoming surveys, such as the Hyper Suprime-Cam (HSC), the Dark Energy Survey (DES), the Kilo Degree Survey (KiDS), the Large Synoptics Survey Telescope (LSST), the Wide-Field Infrared Survey Telescope (WFIRST) and the Euclid mission.

\section*{Acknowledgements}

We thank Juan Garcia-Bellido for the final revision of the paper. 
The MICE simulations have been developed at the MareNostrum supercomputer (BSC-CNS) thanks to grants AECT-2006-2-0011 through AECT-2015-1-0013. Data products have been stored at the Port d'Informació Científica (PIC), and distributed through the CosmoHub webportal (cosmohub.pic.es). Funding for this project was partially provided by the Spanish Ministerio de Ciencia e Innovacion (MICINN), projects 200850I176, AYA2009-13936, AYA2012-39620, AYA2013-44327, ESP2013-48274, ESP2014-58384 , Consolider-Ingenio CSD2007- 00060, research project 2009-SGR-1398 from Generalitat de Catalunya, and the Ramon y Cajal MICINN program. 
AP was supported by beca FI and 2009-SGR-1398 from Generalitat de Catalunya and project AYA2012-39620 from MICINN. 
CC, AA and AR are supported by the Swiss National Science Foundation grants 200021-149442 and 200021-143906.
PF is funded by MINECO, project ESP2013-48274-C3-1-P.

\bibliography{aamnem99,biblist}

\label{lastpage}

\end{document}